\documentstyle[12pt,epsf]{article}
\catcode`\@=11
\def\numberbysection{\@addtoreset{equation}{section}
\def\theequation{\arabic{section}.\arabic{equation}}}
\def\appendix#1{
  \addtocounter{section}{1}
  \setcounter{equation}{0}
  \renewcommand{\thesection}{\Alph{section}}
  \renewcommand{\theequation}{\Alph{section}.\arabic{equation}}
  \section*{Appendix \thesection\protect\indent #1}
  \addcontentsline{toc}{section}{Appendix \thesection\ \ \ #1}
  }
\renewcommand{\thefootnote}{\fnsymbol{footnote}}
\newlength{\extraspace}
\newlength{\extraspaces}
\newcommand{\be}{\begin{equation}
\addtolength{\abovedisplayskip}{\extraspaces}
\addtolength{\belowdisplayskip}{\extraspaces}
\addtolength{\abovedisplayshortskip}{\extraspace}
\addtolength{\belowdisplayshortskip}{\extraspace}}
\newcommand{\ee}{\end{equation}}
\newcommand{\ba}{\begin{eqnarray}
\addtolength{\abovedisplayskip}{\extraspaces}
\addtolength{\belowdisplayskip}{\extraspaces}
\addtolength{\abovedisplayshortskip}{\extraspace}
\addtolength{\belowdisplayshortskip}{\extraspace}}
\newcommand{\ea}{\end{eqnarray}}
\newcommand{\nonu}{\nonumber}
\newcommand{\tr}{\, {\rm tr} \,}
\newcommand{\e}{\, {\rm e}}
\newcommand{\ie}{{\it i.e.}\ }
\begin{document}
\addtolength{\baselineskip}{.5mm}
\thispagestyle{empty}
%
\begin{flushright}
OU--HET 214 \\
June, 1995 \\
hep-th/9506103 \\
\end{flushright}
\vspace{3mm}
\begin{center}
{\Large{\bf{ GENERATING FUNCTIONS IN TWO\\[2mm]
             DIMENSIONAL QUANTUM GRAVITY}}}
\\[24mm]
{\sc Hiroshi Shirokura}\footnote{A JSPS Research Fellow,\quad
e-mail: siro@funpth.phys.sci.osaka-u.ac.jp} \\[8mm]
{\it Department of Physics, Osaka University \\[3mm]
Toyonaka, Osaka 560, JAPAN} \\[29mm]
%
{\bf ABSTRACT}\\[9mm]
{\parbox{13cm}{\hspace{5mm}
We solve general 1-matrix models without taking the double scaling
limit.
A method of computing generating functions is presented.
We calculate the generating functions for a simple and double torus.
Our method is also applicable to more higher genus.
Each generating function can be expressed by a ``specific heat''
function for sphere.
Universal terms, which survived in the double scaling limit can be
easily picked out from our exact solutions.
We also find that the regular part of the spherical generating function
is at most bilinear in coupling constants of source terms.
}}
\end{center}
\vfill
\newpage
\renewcommand{\thefootnote}{\arabic{footnote}}
\setcounter{section}{0}
\setcounter{equation}{0}
\setcounter{footnote}{0}
\numberbysection
%
%
\section{Introduction}

Matrix models \cite{GRMI,BIZ,BIPZ,BRKA} in the double scaling
limit \cite{GRMI}
enable us to treat two dimensional gravity nonperturbatively.
This special limit is necessary for us to investigate
the universality of matrix models.
Many aspects of universality hidden in these simple models have
been revealed by a lot of people.
Matrix models in the double scaling limit are described by
the string equations, a nonlinear differential equation for the
full specific heat function.
The topological expansion of this specific heat requires a large
cosmological constant $t$.
If one want to define the specific heat for small $t$,
nonperturbative input must be demanded.
This point makes the physical meaning of the double scaling limit
unclear.

Our aim of this article is to solve general 1-matrix models exactly
as random lattice models.
Until now matrix models have been solved in the double scaling
limit or in simple potential cases.
Our solution of general 1-matrix models are exact even when
the matrix size $N$ is finite.
All of a 1-matrix model can be shut in a function.
We find the generating function for each topology when the function
is given.
To study non-universal aspects of 1-matrix models,
we rearrange the notion of the scaling operators \cite{DDK} in
these discrete models.
Using this refined formalism, we can understand a lot about the bulk
contributions and the finite size effects of general 1-matrix models.
Our exact solution for general 1-matrix models are important,
since these are precious examples of exactly solvable lattice
models as well as toy models of gravity.
We can easily pick out contributions that survive in the double
scaling limit from our solutions.
This approach to the universal contributions gives us an
explicit answer to the leading generating functions.
As we can include source terms of the physical observables,
we can understand the correlation function of these observables
more directly than the double scaling limit approach.
We cannot find equations like the string equations in our
formalism, however.
We may rather consider that the double scaling limit approach
and ours are complements each of the other.

Before we show main results, we briefly review known results given
by Bessis, Itzykson and Zuber (BIZ) \cite{BIZ}.
A general 1-matrix model is defined by a integral over
a $N\times N$ hermitian matrix $\Phi$,
\ba
Z_N(\{\lambda\}) & = & \int\!\! D\Phi\,\e^{-NS[\Phi,\{\lambda\}]}
                       \,,\nonu \\
S[\Phi,\{\lambda\}] & = &
\tr\left[
     \frac{1}{2}\Phi^2+\sum_{p=1}^\infty
     \lambda_p\Phi^{2p+2}
   \right]\,.
\label{MatrixAction}
\ea
We always fix the coefficient of $\tr\Phi^2$ to $1/2$
in the action.
If we rearrange and fine tune the coupling coefficients
$\{\lambda\}$, the system behaves as one of the Kazakov series
with some source terms.

Using the orthogonal polynomial technique, the generating function
of this model is
\footnote{Our definition for $E_N$ is slightly different from
Eq.(4.3) of \cite{BIZ}. Their sign of $e^{(h)}$ is also opposite
to ours. Take care when comparing their results with ours.}
\ba
E_N(\{\lambda\}) & \equiv &
\frac{1}{N^2} \log \frac{Z_N(\{\lambda\})}{Z_N(0)} \nonu \\
& = &
 \frac{1}{N}\sum_{k=1}^N
 \left(1-\frac{k}{N}\right)\log\frac{R_k(\{\lambda\})}{k}
 + \frac{1}{N}\log\frac{h_0(\{\lambda\})}{h_0(0)}\,.
\label{EN}
\ea
If we denote the orthogonal polynomial of degree $k$ as $P_k(\mu)$,
$R_k(\{\lambda\})$ is defined by a recursion relation for the
polynomials,
\be
\mu P_k(\mu) = P_{k+1}(\mu)+R_k P_{k-1}(\mu)\,.
\label{RecP}
\ee
$h_0$ in (\ref{EN}) is defined as,
\be
h_0(\{\lambda\},N) = \int_{-\infty}^\infty\!\!d\mu\e^{-V(\mu, N)}\,,
\ee
where
\be
V(\mu, N) =
\frac{1}{2}\mu^2
+\sum_{p=1}^\infty\frac{\lambda_p}{N^p}\mu^{2p+2}\,.
\ee
The most characteristic point of this model is that
the large-$N$ expansion of $E_N$ is a genus expansion
\be
E_N = \sum_{h=0}^\infty N^{-2h}\,e^{(h)}\,,
\label{TopEx}
\ee
where $e^{(h)}$ is a contribution from genus $h$.

The authors of \cite{BIZ} expanded
$E_N$ using the Euler-Maclaurin formula and obtained $e^{(h)}$
formally.
Let us introduce some notations to show their result compactly.
It is convenient to note $\epsilon = 1/N,\ x = k/N$
and introduce $r_\epsilon$ and its $\epsilon$-expansion,
\be
r_\epsilon(x,\{\lambda\}) \equiv R_k(\{\lambda\})/N
            = \sum_{s=0}^\infty \epsilon^{2s}r_{2s}(x,\{\lambda\})
\,.
\ee
$r_\epsilon$ satisfies a recursion relation derived from
(\ref{RecP}),
\be
x = r_\epsilon(x,\{\lambda\})
    \left[
     1+\sum_{p=1}^\infty 2(p+1)\lambda_p\sum_{\mbox{\small paths}}
     r_\epsilon(x+s_1\epsilon,\{\lambda\})
     \cdots
     r_\epsilon(x+s_p\epsilon,\{\lambda\})
    \right]\,,
\label{Recr}
\ee
where $\sum_{\mbox{\small paths}}$ means a sum over the ${}_{2p+1}C_p$
paths along a staircase depicted in Fig.1 and integers
$s_i,\ i=1,2,\ldots,p\,$ are the height of the stair when $i$-th
descending down.
For example, we can associate
$\{s_1,s_2,s_3,s_4,s_5\}=\{-1,-1,-2,0,1\}$ for the path drawn in
Fig.1.
%
%
\begin{center}
\leavevmode
\epsfysize=10cm \epsfbox{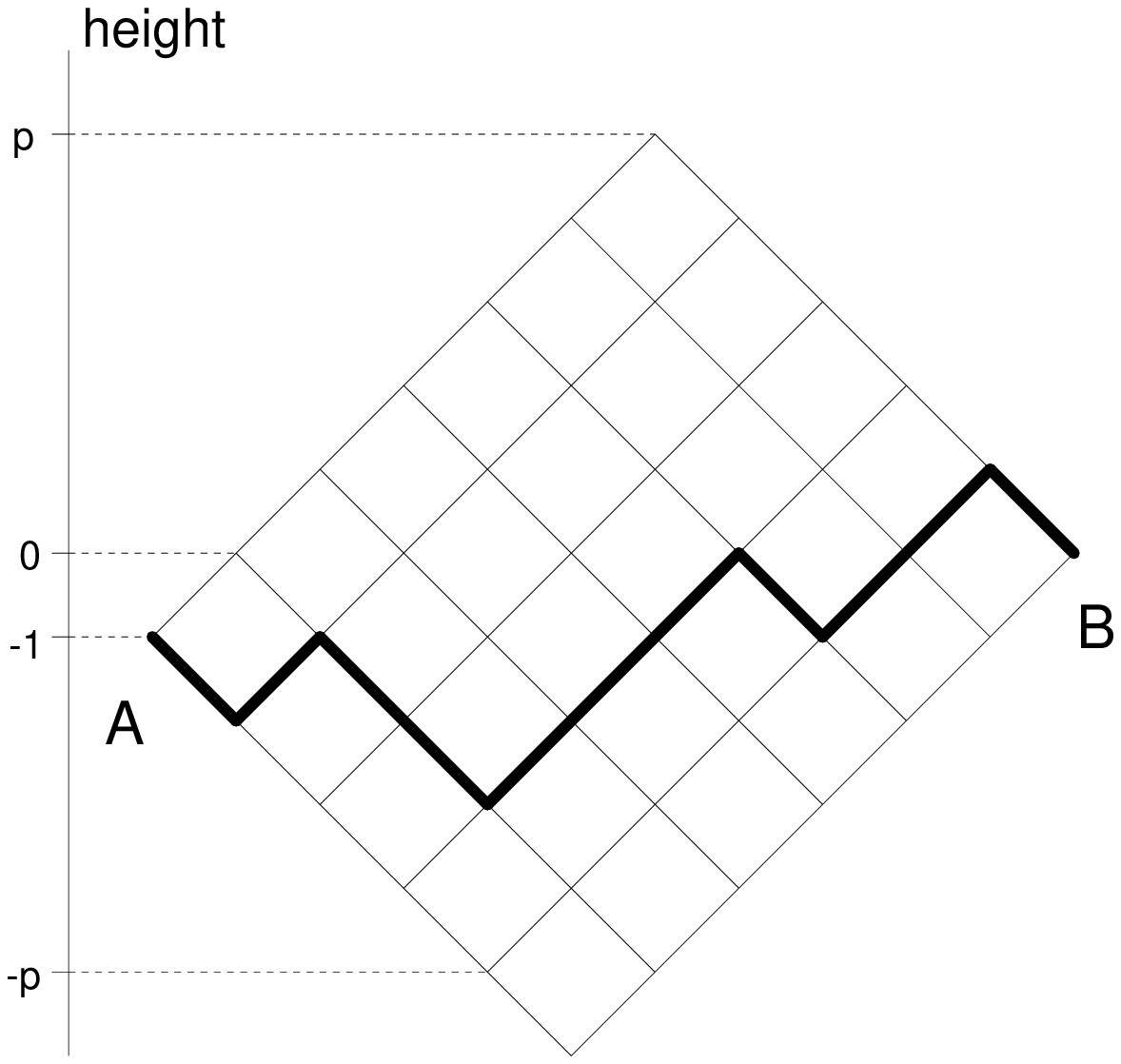}
\end{center}
\begin{center}
{\parbox{13cm}{
\small
Fig.1: An example of path from point A to point B.
There are ${}_{2p+1}C_p$ paths which connect two diagonal points
of a lattice of size $p\times(p+1)$.
In this example, $p=5$.
We assign a set of $p$ integers for each of the paths.
}}
\end{center}
Expanding both sides of this recursion relation in $\epsilon^2$
and comparing the coefficients of $\epsilon^{2s}$,
we can solve $r_{2s}$.
If we take a limit $\epsilon\rightarrow 0$ in (\ref{Recr}),
we get a function $W(r_0)$ which have all information of the
model,
\be
x = r_0 \left[
          1+\sum_{p=1}^\infty \frac{(2p+2)!}{p!(p+1)!}\lambda_p r_0^p
        \right]
  \equiv W(r_0)\,.
\label{StringEq}
\ee
Note that
\[
W(0)=0,\,W'(0)=1,\,\lim_{r_0\rightarrow 0}\frac{W(r_0)}{r_0}=1\,.
\]
One of our goal is to understand $e^{(h)}$ for given $W(r_0)$.
We introduce two new notations.
The first is defined by a large-$N$ expansion
\be
\log\frac{r_\epsilon(x,\{\lambda\})}{x}\equiv
\sum_{s=0}^\infty\frac{1}{N^{2s}}\tilde{r}_{2s}
(x,\{\lambda\})\,.
\label{tilder}
\ee
We show the first few of them:
\ba
\tilde{r}_0 & = & \log\frac{r_0}{x}\,,\quad
\tilde{r}_2   =   \frac{r_2}{r_0}\,, \nonu \\
\tilde{r}_4 & = & \frac{r_4}{r_0}-\frac{1}{2}
                  \left( \frac{r_2}{r_0} \right)^2\,.
\ea
In our formalism, these quantities play an important role.
The next is associated with the last term of the right hand side of
(\ref{EN}) \cite{BIZ},
\be
\frac{1}{2N}
\left\{
  \log\frac{h_0(\{\lambda\}, N)}{h_0(0)}
 -\log\frac{h_0(\{\lambda\},-N)}{h_0(0)}
\right\}
\equiv
\sum_{s=1}^\infty\frac{1}{N^{2s}}K_{2s}(\{\lambda\})\,.
\ee
Since $\lambda_p$ is rewritten by a derivative of $W(r_0)$
at $r_0=0$,
\ba
h_0(\{\lambda\}, N) & = &
\int_{-\infty}^\infty\!\!d\mu\,\e^{-V(\mu, N)}
\nonu \\
& = &
\int_{-\infty}^\infty\!\!d\mu\,\exp
  \left[
     -\frac{1}{2}\mu^2
     -\sum_{p=1}^\infty\frac{p!}{(2p+2)!}
      \frac{W^{(p+1)}(0)}{N^p}\mu^{2p+2}
  \right]\,.
\ea
The first few of them are:
\ba
K_0 & \equiv & 0, \quad K_2 = -\frac{1}{8}W''(0)\,, \nonu \\
K_4 &   =    & -\frac{1}{192}
                \left[
                 22\{W''(0)\}^3-20W''(0)W'''(0)+3W^{(4)}(0)
                \right]\,.
\ea
Then the contribution from genus $h$ is written as
\ba
e^{(h)} & = &
\int_0^1\!\!dx(1-x)\tilde{r}_{2h}(x,\{\lambda\})
+ K_{2h}(\{\lambda\}) \nonu \\
&   &
-\sum_{s=1}^h (-1)^s \frac{B_s}{(2s)!}
 \left.
  \left\{
   (1-x)\tilde{r}_{2(h-s)}(x,\{\lambda\})
  \right\}^{(2s-1)}
 \right|_0^1\,,
\label{GeneralGF}
\ea
where $B_s$ are the Bernoulli numbers
\be
B_1 = \frac{1}{6},\quad B_2 = \frac{1}{30},\quad
B_3 = \frac{1}{42}\,, \ldots \,.
\ee
In (\ref{GeneralGF}) the superscript $(2s-1)$ stands for the
$(2s-1)$-th derivative
with respect to $x$ and the symbol $|^1_0$ means taking the
difference between the values of the function at $x=1$ and $x=0$.
We can directly solve the recursion relation for $r_{2s}$ and
calculate the generating functions from
(\ref{GeneralGF}) when $W(r_0)$ is simple.
It is, however, a troublesome task to find $e^{(h)}$ directly
from (\ref{GeneralGF}) even when $W(r_0)$ is quadratic.
There are roughly two way to advance:
to take the double scaling limit and drop all non-universal
information, or to solve (\ref{GeneralGF}) for general cases
more explicitly.
Our method achieves the latter.

%
%
The main outcome of this article is the generating functions for
a simple and double torus.
The answer for sphere has already been given in \cite{BIZ}.
If we use $a^2$ such that $W(a^2) = 1$,
the answer for $e^{(1)}$ is very simple,
\be
e^{(1)} = -\frac{1}{12}\log\{a^2W'(a^2)\}\,.
\ee
On the other hand $e^{(2)}$ is a little complex,
\ba
e^{(2)} & = &
\frac{a^2
      \left[
       28\{W''(a^2)\}^3-29W'(a^2)W''(a^2)W'''(a^2)
       +5\{W'(a^2)\}^2W^{(4)}(a^2)
      \right]}{1440\{W'(a^2)\}^5}
\nonu \\
&   &
-\frac{9\{W''(a^2)\}^2-4W'(a^2)W'''(a^2)}
      {2880\{W'(a^2)\}^4}
\nonu \\
&   &
+\frac{1}{240}
-\frac{1}{240}\frac{W''(a^2)}{a^2\{W'(a^2)\}^3}
-\frac{1}{240}\frac{1}{a^4\{W''(a^2)\}^2}\,.
\ea
Though we do not compute $e^{(h)}$ for $h\geq 3$ this time,
there are no obstacles.
They will be more complex than $e^{(2)}$.
These explicit answers for $e^{(h)}$ allows us to compute
exact correlation functions of scaling operators.

%
%
The outline of this paper is as follows.
In section 2 we make a systematic study on the representations of
scaling operators by the matrix $\Phi$.
We write down the polynomial $W(r)$ in which all information about
the $k$-th model with source terms are packed.
The formulation in this section shows its ability when
we study non-universal aspects of matrix models as well as
the universality.
In section 3 we compute the spherical generating function
for the general setting of section 2
and show that the regular part is naturally separated from the
result.
Section 4 is the main part of this paper.
We give here a method to perform the program of BIZ thoroughly for
general $W(r)$.
It is essential to understand weighted summations over paths appeared
in the formulation of BIZ.
We solve them using generating functions for these sums.
As an application of the result of section 4, we compute the
correlation functions of scaling operators for a simple and double torus
in section 5.
We also study the leading singular terms which survive in the
double scaling limit.
We discuss advantages and disadvantages of our method in section 6.
In appendix A we prove that a generating function of
a special weighted sums over paths is a $q$-deformed combination.
Appendix B is devoted to another generating function for sums over
the paths.

%
\section{Fundamental Representations for Scaling Operators}

The scaling operators in 1-matrix models have simple scaling
properties \cite{GRMI}.
This scaling properties are specified by the string susceptibility
$\gamma$ and gravitational dimensions of scaling operators.
The $k$-th criticality is realized by a fine tuning of the
parameters $\{\lambda\}$ in (\ref{MatrixAction}).
The string susceptibility for the $k$-th model is $-1/k$.
It is well-known that the gravitational dimension of the scaling
operator in the $k$-th model is a multiple of
the absolute value of the string susceptibility
$|\gamma|=1/k$.
We can express these objects by the matrix $\Phi$.
There are some ambiguities in how to represent these scaling
operators with respect to $\Phi$.
Generally, a scaling operator with gravitational dimension $d$
is written in a form of a linear combination of the following
infinite set of ``fundamental representations'',
\be
\rho_d^M[\Phi],\quad d=\frac{n}{k}\, \quad n,M = 0,1,2,\ldots\ .
\ee
The definition of these fundamental representations is simply given
by a generation rule
%
%
\ba
\label{GenerationRule}
\rho_{n/k}^M[\Phi] & = &  \rho_{(n-1)/k}^M[\Phi]
                         -\rho_{(n-1)/k}^{M+1}[\Phi]\,, \\
\rho_0^M[\Phi]     & = & \frac{(M+1)!(M+2)!}{(2M+4)!}k^{-M-2}
                         \tr\Phi^{2M+4}\,.
\label{Puncture}
\ea
This generation rule (\ref{GenerationRule}) means that each
fundamental representation is a linear combination of puncture
operators (\ref{Puncture})
\be
\rho_{n/k}^M[\Phi] = \sum_{p=0}^n
                      \left(
                       \begin{array}{c}
                        n \\ p
                       \end{array}
                      \right)
                     (-1)^p\rho_0^{M+p}[\Phi]\,,
\label{ExpansionByPuncture}
\ee
where
\[
\left(
 \begin{array}{c}
  n \\ p
 \end{array}
\right)
\equiv {}_nC_p = \frac{n!}{p!(n-p)!}\,.
\]
The number of puncture operators in (\ref{ExpansionByPuncture})
corresponds to the gravitational dimension $d=n/k$ of the scaling
operator.
The index $M$ specifies the puncture operator with the lowest degree
when expanded by the puncture operators.
In Section 5 we will show that the scaling operators with different
$M$ degenerate in the double scaling limit.
This means that the index $M$ is non-universal.
For example, $\rho_1^1$ in pure gravity $(k=2)$ is
\ba
\rho_1^1[\Phi] & = &  \rho_0^1[\Phi]-2\rho_0^2[\Phi]
                     +\rho_0^3[\Phi] \nonu \\
               & = &  \frac{1}{480}  \tr\Phi^6
                     -\frac{1}{2240} \tr\Phi^8
                     +\frac{1}{40320}\tr\Phi^{10} \nonu \\
               & = &  \frac{1}{40320}\tr
                      \left(
                       \Phi^{10}-18\Phi^8+84\Phi^6
                      \right)\,.
\ea

Using these fundamental representations, we can formulate a general
model, which is fine tuned in the $k$-th critical point and has
$(m+1)$ source terms.
We reserve the $0$-th entry as the cosmological constant term.
The action for this model is
\be
S[\Phi] = S_{\mbox{\small cr}}[\Phi]
          + \sum_{i=0}^m\tau_i\rho_{n_i/k}^{M_i}[\Phi]\,,
\label{GeneralAction}
\ee
where the fine tuned part $S_{\mbox{\small cr}}[\Phi]$
is formally written by a special scaling operator
with negative $M$ and dimension one,
\ba
S_{\mbox{\small cr}}[\Phi]
 & = & N - \rho_1^{-2}[\Phi] \nonu \\
 & = & \tr\left[
           \frac{1}{2}\Phi^2-\sum_{p=2}^k
            \left(
             \begin{array}{c}
              k \\ p
             \end{array}
            \right)
           (-1)^p\frac{(p-1)!p!}{(2p)!}k^{-p}\Phi^{2p}
          \right]\,.
\ea
The existence of this special scaling operator prevents us
define fundamental representations with negative $M$.
These irregular representations change the coefficient of $\tr\Phi^2$
in (\ref{GeneralAction}).
We can avoid these representations by redefinition of $\Phi$ and
the coupling parameters $\tau_i$.
If we formally define that
$n_{-1}\equiv k,\ M_{-1}\equiv -2,\ \tau_{-1}\equiv -1$,
(\ref{GeneralAction}) becomes simpler,
\be
S[\Phi] = N + \sum_{i=-1}^m\tau_i\rho_{n_i/k}^{M_i}[\Phi]\,.
\ee
We can translate this action in the form of $W(r)$ appeared in the
previous section,
\be
W(r) = 1 - \sum_{i=-1}^m
           \left(\frac{r}{k}\right)^{2+M_i}
           \left(1-\frac{r}{k}\right)^{n_i}
           \tau_i\,.
\label{GeneralW}
\ee
This fact justifies our definition of the fundamental representations.
The definition (\ref{GenerationRule}) has a simple meaning:
each scaling operator is constructed by cancelling the leading
singularity of the two distinct scaling operators with dimension
subtracted by $|\gamma|=1/k$ from the original one.
Our setting of the 1-matrix models given by (\ref{GeneralW})
is very useful in later analysis of generating functions.


Strictly speaking, the definition for the fundamental
representations depends on the choice of the fine tuned part
of the action $S_{\mbox{\small cr}}$.
There is a degree of freedom for deformation of the fundamental
representation $\rho_{n/k}^M$.
This degree of freedom is related to the choice of
$S_{\mbox{\small cr}}$.
The modified generation rule is similar to the previous one
\ba
\rho_{n/k}^M[\Phi,b] & = &
 \rho_{(n-1)/k}^M[\Phi,b]-\rho_{(n-1)/k}^{M+1}[\Phi,b]\,, \nonu \\
\rho_0^M[\Phi,b]     & = &
 \frac{(M+1)!(M+2)!}{(2M+4)!}(k+b)^{-M-2}\tr\Phi^{2M+4}\,.
\label{ModGenerationRule}
\ea
The deformation parameter $b$ must be greater than $-k$.
When $b=0$, (\ref{ModGenerationRule}) gets back to the generation
rule (\ref{GenerationRule}).
These deformed representations correspond to a fine tuned
action
\be
S_{\mbox{\small cr}}[\Phi,b] =
 N - \rho_1^{-2}[\Phi,b] + b\rho_1^{-1}[\Phi,b]\,.
\label{DeformedCriticalAction}
\ee
This additive term is a special marginal scaling operators as
before.
We assign $M_{-2}\equiv -1,\ n_{-2}\equiv k,\ \tau_{-2}\equiv b$,
for the additive term in (\ref{DeformedCriticalAction}).
Then the whole action is translated to
\ba
W(r) & = & 1-\left(1-b\frac{r}{k+b}\right)
           \left(1-\frac{r}{k+b}\right)^k
           +\sum_{i=0}^m\left(\frac{r}{k+b}\right)^{M_i+2}
           \left(1-\frac{r}{k+b}\right)^{n_i}\tau_i \nonu \\
     & = & 1 - \sum_{i=-2}^m\left(\frac{r}{k+b}\right)^{M_i+2}
           \left(1-\frac{r}{k+b}\right)^{n_i}\tau_i\,.
\ea
We can easily see that
$\rho_{n/k}^M[\Phi,1] = \rho_{n/(k+1)}^M[\Phi,0]$.
Therefore the parameter $b$ connects the $k$-th criticality with
the $(k+1)$-th criticality of 1-matrix models.
We can add ordinary marginal operators, \ie,
operators with gravitational dimension one and non-negative $M$
to the critical action $S_{\mbox{\small cr}}$.
These terms, however, do not cause further modification in the
fundamental representations.
Since a coupling constant $\tau_i$ do not scale when its dimension
is one, we should regard these marginal operators as source terms.
In this point of view, the most general fine tuned action consists of
marginal operators with $M=-2$ and $M=-1$.
The special marginal operator $\rho_1^{-2}$ is always necessary
to cancel $N$ in the fine tuned action.
In section 5, we will show that this parameter $b$ (if $b\neq 1$)
is non-universal as the index $M$.
{}From now on, we will mainly discuss the case of $b=0$ for simplicity.

We conclude this section by classifying quantities which characterize
general 1-matrix models.

\newlength{\listlength}
\settowidth{\listlength}{$M_i$}
\begin{enumerate}
\item Universal quantities.
\begin{list}{}{%
\setlength{\leftmargin}{\listlength}
\setlength{\labelsep}{3em}
\addtolength{\leftmargin}{\labelsep}
\setlength{\labelwidth}{\listlength}
\renewcommand{\makelabel}{}
}
\item[$k$\hfill]
specifies the criticality of the system.
The string susceptibility $\gamma$ depends only on this integer,
$\gamma=-1/k$.
\item[$n_i$\hfill]
is related to a gravitational dimension of a scaling operator,
$d_i=n_i/k$.
\end{list}
\item Non-universal quantities.
\begin{list}{}{%
\setlength{\leftmargin}{\listlength}
\setlength{\labelsep}{3em}
\addtolength{\leftmargin}{\labelsep}
\setlength{\labelwidth}{\listlength}
\renewcommand{\makelabel}{}
}
\item[$b$\hfill]
modifies the fine tuned part of the action.
$b$ also connects adjacent criticalities.
\item[$M_i$\hfill]
shows degrees of freedom in representing scaling operators
with the matrix $\Phi$.
A scaling operator is constructed by taking a difference of the two
scaling operators with adjacent values of $M_i$ and with
gravitational dimension lower than the original one by $1/k$.
\end{list}
\end{enumerate}

%
\section{Non-universal Part in Generating Function for Sphere}

The planar contribution of the generating function has been argued
since early times \cite{GRMI,BIZ,BIPZ}.
In this section we represent the scaling nature of spherical part
without introducing any scaling units such as the cosmological
constant.

The spherical generation function $e^{(0)}$ is written by
the function $W(r)$ as
\ba
e^{(0)} & = & \int_0^1\!\!dx(1-x)\log\frac{r_0}{x} \nonu \\
        & = & -\int_0^{a^2}\!\!drW'(r)[1-W(r)]\log
              \left\{ \frac{W(r)}{r} \right\}\,.
\ea
Integrating by parts, we can rewrite this in the following form,
\be
e^{(0)} = \frac{1}{2}\log a^2
         +\frac{1}{2}\int_0^{a^2}\!\!dr\,r\bar{W}'(r)[2-W(r)]\,,
\label{RewrittenE0}
\ee
where $\bar{W}$ is
\be
\bar{W}(r)\equiv\frac{W(r)}{r}\,.
\ee
If we substitute the most general form for $W(r)$
\be
W(r) =
r + \sum_{p=1}^\infty\frac{(2p+2)!}{p!(p+1)!}\lambda_p r^{p+1}
\ee
to (\ref{RewrittenE0}) and eliminate $\lambda_1$ using
$W(a^2) = 1$, we have
\ba
e^{(0)}
 & = &
 \frac{1}{2}\log a^2 + \frac{1}{24}(1-a^2)(9-a^2) \nonu \\
 &   &
 + \sum_{p=2}^\infty
   \left[
     \frac{(p-1)(p+5)(2p+1)!}{2(p+3)(p+1)!p!} a^{2p+2}
    -\frac{(p-1)(p+6)(2p+2)!}{12(p+3)!p!} a^{2p+4}
   \right] \lambda_p \nonu \\
 &   &
 + \sum_{p=2}^\infty\sum_{q=2}^\infty
   \frac{(p-1)(q-1)(p+q+6)(2p+2)!(2q+2)!}
        {8(p+3)(q+3)(p+q+2)(p+1)!p!(q+1)!q!} a^{2p+2q+2}
   \lambda_p\lambda_q\,.
\ea
Here $\lambda_1$ is nothing but the cosmological constant.
When $\lambda_p = 0,\ (p=2,3,\ldots)$, this result is the same as
that of \cite{BIZ}.
This form, however, is not fit for understanding the scaling property
of the generating function.
We must fine tune to the $k$-th criticality and introduce
source terms to understand universality hidden in matrix models.
The setting of the previous section,
\be
W(r) = 1 - \left( 1-\frac{r}{k} \right)^k
         + \sum_{i=0}^m \left(\frac{r}{k}\right)^{M_i+2}
           \left( 1-\frac{r}{k} \right)^{n_i} \tau_i
\ee
is suitable for our aim.
Substituting this function in (\ref{RewrittenE0}), we visualize
the scaling behavior of $e^{(0)}$ roughly.
The integrand in (\ref{RewrittenE0}) is described by
powers of $1-r/k$.
The $\{\tau\}$-independent terms in $e^{(0)}$ are
\ba
\lefteqn{
  \frac{1}{2}\log a^2
 +\frac{1}{2}\int_0^{a^2}\!\!dr
 \left[
  -\frac{1}{k}\sum_{p=0}^{k-1}\left( 1-\frac{r}{k} \right)^p
  +\left( 1-\frac{r}{k} \right)^{k-1}
 \right]
 \left[
  1 + \left( 1-\frac{r}{k} \right)^k
 \right]
} \nonu \\
& = &
 \left(
  \frac{1}{2}\log k+\frac{3}{4}-\frac{1}{2}\sum_{p=1}^{2k}\frac{1}{p}
 \right)
 -\frac{1}{2}\sum_{p=0}^\infty\frac{1}{2k+1+p}
  \left( 1-\frac{a^2}{k} \right)^{2k+1+p}
\nonu \\
&   &
 -\frac{1}{2} \left( 1-\frac{a^2}{k} \right)^k
 -\frac{1}{4} \left( 1-\frac{a^2}{k} \right)^{2k}\,.
\ea
Picking up linear terms in $\{\tau\}$, we find
\ba
\lefteqn{
 \frac{1}{2}\sum_{i=0}^m\tau_i\int_0^{a^2}\!\!dr
 \left[
  \left\{
   \frac{1}{r}\left(
               1 - \left( 1-\frac{r}{k} \right)^k
              \right)
   -\left( 1-\frac{r}{k} \right)^{k-1}
  \right\}
  \left(\frac{r}{k}\right)^{M_i+2}
  \left( 1-\frac{r}{k} \right)^{n_i}
 \right.
} \nonu \\
 &   &
 \left.
  + \left(\frac{r}{k}\right)^{M_i+1}
   \left\{
     \frac{M_i+n_i+1}{k}
    \left( 1-\frac{r}{k} \right)^{n_i}
    -\frac{n_i}{k}
    \left( 1-\frac{r}{k} \right)^{n_i-1}
   \right\}
   \left\{
    1+ \left( 1-\frac{r}{k} \right)^k
   \right\}
 \right] \nonu \\
 & = &
 \sum_{i=0}^m\tau_i
 \left[
  \sum_{p=0}^{M_i+1}\frac{(-1)^p}{k+n_i+1+p}
  \left(
   \begin{array}{c}
    M_i+1 \\ p
   \end{array}
  \right)
  \left\{
   \left( 1-\frac{a^2}{k} \right)^{k+n_i+1+p} - 1
  \right\}
 \right. \nonu \\
 &   &
 \left.
 +\frac{1}{2}\left(\frac{a^2}{k}\right)^{M_i+2}
  \left( 1-\frac{a^2}{k} \right)^{n_i}
 +\frac{1}{2}\left(\frac{a^2}{k}\right)^{M_i+2}
  \left( 1-\frac{a^2}{k} \right)^{k+n_i}
 \right]\,.
\label{LinearInTau}
\ea
To derive the second and the third term of the most left hand side
of (\ref{LinearInTau}), we used a property of binomial coefficients
such as
\be
\left(
 \begin{array}{c}
  M_i+1 \\ p
 \end{array}
\right)
(M_i+n_i+2)+
\left(
 \begin{array}{c}
  M_i+1 \\ p-1
 \end{array}
\right)
n_i =
\left(
 \begin{array}{c}
  M_i+2 \\ p
 \end{array}
\right)
(M_i+n_i+2-p)\,.
\ee
The bilinear terms in $\{\tau\}$ are
\ba
\lefteqn{
 -\frac{1}{4}\sum_{i=0}^m\sum_{j=0}^m\tau_i\tau_j\int_0^{a^2}\!\!dr
 \left(\frac{r}{k}\right)^{M_i+M_j+3}
} \nonu \\
& \times &
 \left[
  \left( 1-\frac{r}{k} \right)^{n_i}
  \left\{
    \frac{M_j+n_j+1}{k} \left( 1-\frac{r}{k} \right)^{n_j}
   -\frac{n_j}{k}       \left( 1-\frac{r}{k} \right)^{n_j-1}
  \right\}
  + (i\leftrightarrow j)
 \right]
\nonu \\
 & = &
 \sum_{i=0}^m\sum_{j=0}^m\tau_i\tau_j
 \left[
  -\frac{1}{2}\sum_{p=0}^{M_i+M_j+3}
  \frac{(-1)^p}{n_i+n_j+1+p}
  \left(
   \begin{array}{c}
    M_i+M_j+3 \\ p
   \end{array}
  \right)
 \right.
\nonu \\
&   &
\times
 \left.
  \left\{
   \left( 1-\frac{a^2}{k} \right)^{n_i+n_j+1+p} - 1
  \right\}
  -\frac{1}{4}\left(\frac{a^2}{k}\right)^{M_i+M_j+4}
  \left( 1-\frac{a^2}{k} \right)^{n_i+n_j}
 \right].
\ea
Here we again used the above nature of binomial coefficients.
Introducing the following rules for binomial coefficients
\be
\left(
 \begin{array}{c}
  -1 \\ p
 \end{array}
\right)
\equiv (-1)^p\quad(\mbox{for}\ p\ge 0),\qquad
\left(
 \begin{array}{c}
  n \\ p
 \end{array}
\right)
\equiv 0\quad(\mbox{for}\ p>n)\,,
\ee
and using $W(a^2)=1$,
we can assemble all contributions in a compact form,
\ba
e^{(0)}
 &\!\! = \!\!& \frac{1}{2}\log k+\frac{3}{4}
      -\frac{1}{2}\sum_{p=1}^\infty\frac{1}{p} \nonu \\
 &\!\!   \!\!\!&
 -\frac{1}{2}\sum_{i=-1}^m\sum_{j=-1}^m\tau_i\tau_j
 \sum_{p=0}^\infty\frac{(-1)^p}{n_i+n_j+1+p}
 \left(
  \begin{array}{c}
   M_i+M_j+3 \\ p
  \end{array}
 \right) \nonu \\
 &\!\!   \!\!\!& \times
 \left[ \left(1-\frac{a^2}{k}\right)^{n_i+n_j+1+p}-1 \right]\,,
\label{ScalingFormE0}
\ea
where negatively labeled quantities
This final result naturally separates into the regular part
and the rest,
\be
e^{(0)} = e^{(0)}_{\mbox{\small bulk}}+e^{(0)}_{\mbox{\small sing}}\,,
\label{bulksing}
\ee
where
\ba
e^{(0)}_{\mbox{\small bulk}} & = &
 \frac{1}{2}\log k+\frac{3}{4}-\frac{1}{2}\sum_{p=1}^{2k}\frac{1}{p}
 -\sum_{i=0}^m\tau_i\sum_{p=0}^{M_i+1}
 \frac{(-1)^p}{n_i+k+1+p}
 \left(
  \begin{array}{c}
   M_i+1 \\ p
  \end{array}
 \right) \nonu \\
 &   &
 +\frac{1}{2}\sum_{i=0}^m\sum_{j=0}^m\tau_i\tau_j
 \sum_{p=0}^{M_i+M_j+3}\frac{(-1)^p}{n_i+n_j+1+p}
 \left(
  \begin{array}{c}
   M_i+M_j+3 \\ p
  \end{array}
 \right) \nonu \\
 & = &
 \frac{1}{2}\log k+\frac{3}{4}-\frac{1}{2}\sum_{p=1}^{2k}\frac{1}{p}
 -\sum_{i=0}^m\frac{(M_i+1)!(n_i+k)!}{(M_i+n_i+k+2)!}\tau_i \nonu \\
 &   &
 +\frac{1}{2}\sum_{i=0}^m\sum_{j=0}^m
 \frac{(M_i+M_j+3)!(n_i+n_j)!}{M_i+M_j+n_i+n_j+4)!}\tau_i\tau_j
\ea
and
\ba
e^{(0)}_{\mbox{\small sing}} & = &
 -\frac{1}{2} \sum_{i=-1}^m\sum_{j=-1}^m\tau_i\tau_j
 \sum_{p=0}^\infty\frac{(-1)^p}{n_i+n_j+1+p}
 \left(
  \begin{array}{c}
   M_i+M_j+3 \\ p
  \end{array}
 \right)
\nonu \\
&   &
 \times \left(1-\frac{a^2}{k}\right)^{n_i+n_j+1+p}\,.
\label{SingularPart}
\ea
The regular part $e^{(0)}_{\mbox{\small bulk}}$ has a remarkable
nature.
It is at most bilinear in the coupling constants
$\{\tau\}$ and all the coupling constants are equal in
$e^{(0)}_{\mbox{\small bulk}}$.
This result for the non-universal terms is nontrivial, for
we apt to think that the structure of the non-universal part
changes its form according to the operator content of the model.
This naive estimate is not true, however, and there is a universal
structure even in the non-universal part.
We should consider this universal structure in the bulk part
as one of the characteristic of the matrix models.
Though we have no proof, it is natural to think that there is
the same structure in the non-universal part even in 2-matrix
models.
On the other hand we regard $e^{(0)}_{\mbox{\small sing}}$ as
the singular part, because the factor $1-a^2/k$ becomes singular
with respective to a scaling unit which will be introduced later.
Though we have not yet introduced such a scaling unit,
we can see a sign of scaling nature through the factor
$1-a^2/k$.
We should note that there is a simple relation between
$e^{(0)}_{\mbox{\small bulk}}$ and $e^{(0)}_{\mbox{\small sing}}$.
If we replace $a^2$ with $0$ in $e^{(0)}_{\mbox{\small sing}}$,
it equals to $-e^{(0)}_{\mbox{\small bulk}}$ up to
$\{\tau\}$-independent constant (infinite) value.

Our result for the bulk part is crucial in matrix models
modified by trace-squared terms \cite{DDSW,KLEHASHI,SIRO}.
The effective action for baby universes associated with scaling
operators depends deeply on this bulk structure of unmodified
models.
Our consequence shows that the Liouville dressing $\beta$
for a source term always changes its branch when the coupling
constant of a trace-squared term associated with the source is
fine tuned.
This also denies the existence of an upper bound for
the gravitational dimension of scaling operators \cite{SIRO}.


When we deform the system by $b$ as in the section 2, the generating
function is compactly given in the following form
\ba
e^{(0)}(b)
 &\!\! = \!\!&
 \frac{1}{2}\log(k+b)+\frac{3}{4}
 -\frac{1}{2}\sum_{p=1}^\infty\frac{1}{p} \nonu \\
 &\!\!   \!\!\!&
 -\frac{1}{2}\sum_{i=-2}^m\sum_{j=-2}^m\tau_i\tau_j
 \sum_{p=0}^\infty\frac{(-1)^p}{n_i+n_j+1+p}
 \left(
  \begin{array}{c}
   M_i+M_j+3 \\ p
  \end{array}
 \right)
\nonu \\
 &\!\!   \!\!&
\times
 \left[ \left(1-\frac{a^2}{k+b}\right)^{n_i+n_j+1+p}-1 \right]\,,
\ea
where we used notations of section 2 for negatively labeled
quantities.
It is important that the structure of the spherical generating
function (\ref{bulksing}) is kept even when nonzero $b$.
One can easily check that when $b=1$, the system behaves as
the $(k+1)$-th Kazakov series,
\ba
e^{(0)}(b=1)
 &\!\! = \!\!& \frac{1}{2}\log (k+1)+\frac{3}{4}
      -\frac{1}{2}\sum_{p=1}^\infty\frac{1}{p} \nonu \\
 &\!\!   \!\!\!&
 -\frac{1}{2}\sum_{i=-1}^m\sum_{j=-1}^m\tau_i\tau_j
 \sum_{p=0}^\infty\frac{(-1)^p}{n_i+n_j+1+p}
 \left(
  \begin{array}{c}
   M_i+M_j+3 \\ p
  \end{array}
 \right) \nonu \\
 &\!\!   \!\!\!& \times
 \left[ \left(1-\frac{a^2}{k+1}\right)^{n_i+n_j+1+p}-1 \right]\,,
\ea
where $n_{-1}=k+1$ instead of $k$.

%
\section{Generating Functions for Higher Genus}

In this section we compute generating functions for higher genus
with respective to the function $W(r)$ in which all the information
of the model are packed.
We must overcome a difficulty in expressing the quantity
$r_{2s}$ appeared in \cite{BIZ} in terms of $W(r_0)$ to accomplish
this task.
We realize this with the aid of a $q$-deformed combination and
a generating function of summations over paths connecting two
fixed points (See Fig.1 in \S 1).

%
\subsection{Torus}

Our starting point for the toric contribution of the generating
function is read from (\ref{GeneralGF})
\be
e^{(1)} = \int_0^1\!\!dx(1-x)\frac{r_2}{r_0}-\frac{1}{8}W''(0)
          \left.
           +\frac{1}{12}
           \left\{
            (1-x)\log\frac{r_0}{x}
           \right\}^{(1)}
          \right|_0^1\,,
\label{BIZTorus}
\ee
where $x$ is nothing but the function $W(r_0)$.
The last term of (\ref{BIZTorus}) can be written only by $a^2$ and
the derivatives of $W(r_0)$ at $r_0=0$.
To show this we define derivatives of $r_0(x)$ with respect to $X$
as $r_0^{(n)}\equiv(d^n/dx^n)r_0$.
If we differentiate both sides of (\ref{StringEq}) with respect to
$x$, we have
\be
r_0^{(1)} = [W'(r_0)]^{-1}\,.
\ee
So we have
\ba
\left\{
 (1-x)\log\frac{r_0}{x}
\right\}^{(1)}
& = &
-\log\frac{r_0}{x} + \frac{1-x}{r_0}r_0^{(1)}-\frac{1-x}{x}
\nonu \\
& = &
-\log\frac{r_0}{W(r_0)} + \frac{1-W(r_0)}{r_0W'(r_0)}
-\frac{1-W(r_0)}{W(r_0)}\,.
\ea
Since $W(a^2)=1$, the difference between the values of the
above function at $x=1$ and $x=0$ is
\be
\left.
 \frac{1}{12}
 \left\{
  (1-x)\log\frac{r_0}{x}
 \right\}^{(1)}
\right|_0^1
= -\frac{1}{12}\log a^2
  -\frac{1}{12}\lim_{r_0\rightarrow 0}
   \left[
    \frac{1-W(r_0)}{r_0W'(r_0)}-\frac{1-W(r_0)}{W(r_0)}
   \right]\,.
\ee
Using the facts like $W(0)=0, W'(0)=1$, we can find the limit value,
\ba
\lim_{r_0\rightarrow 0}
\left[
 \frac{1-W(r_0)}{r_0W'(r_0)}-\frac{1-W(r_0)}{W(r_0)}
\right]
& = &
\lim_{r_0\rightarrow 0}
\left[
 \frac{1}{r_0W'(r_0)}-\frac{1}{W(r_0)-W(0)}
\right] \nonu \\
& = &
\lim_{r_0\rightarrow 0}
\frac{r_0^2W''(0)}{\{r_0W'(0)\}^2}
\left( -1+\frac{1}{2} \right) \nonu \\
& = &
-\frac{W''(0)}{2}\,.
\ea
Thus we have
\be
\left.
 \frac{1}{12}
 \left\{
  (1-x)\log\frac{r_0}{x}
 \right\}^{(1)}
\right|_0^1
= -\frac{1}{12}\log a^2
  +\frac{1}{24}W''(0)\,.
\ee

We now rewrite $r_2/r_0$ in terms of $W(r_0)$.
To do this we solve the recursion relation for $r_\epsilon$
(\ref{Recr}).
We identify the coefficient of $\epsilon^2$ on both sides of
(\ref{Recr}), and find
\ba
0 & = & r_2 + \sum_{p=1}^\infty 2(p+1)\lambda_p
        \left[
         \left(
          \begin{array}{c}
           2p+1 \\ p
          \end{array}
         \right) r_2 r_0^p
        +r_0%
         \left\{
          \left(
           \begin{array}{c}
            2p+1 \\ p
           \end{array}
          \right) pr_2 r_0^{p-1}
         \right.
        \right. \nonu \\
&   &
\left.
 \left.
  +\sum_{\mbox{\small paths}}\sum_{i=1}^p\frac{1}{2}s_i^2r_0^{(2)}r_0^{p-1}
  +\frac{1}{2}\sum_{\mbox{\small paths}}\sum_{i,j}{}'s_is_j
  \left( r_0^{(1)} \right)^2 r_0^{p-2}
 \right\}
\right]\,,
\label{DefR2}
\ea
where $\sum_{i,j}{}'$ means that the case $i=j$ is excluded in the sum.
There are two unknown summations over paths.
We must find their values for given $p$.
It is difficult to guess the values of these sums for general $p$
from the individual values of the sums at small $p$.
We need some clever devices to understand these sums.

Let us introduce a general definition for such sums over the paths.
For a given partition of an even integer $2n$,
\be
2n = 1\cdot a_1+2\cdot a_2+\cdots+(2n)\cdot a_{2n},
\ee
we define a summation over the paths as
\be
I_p(1^{a_1}\cdot 2^{a_2}\cdots(2n)^{a_{2n}}) =
\left[ \prod_{q=1}^{2n}a_q! \right]^{-1}
\sum_{\mbox{\small paths}}
\left[ \prod_{q=1}^{2n}(q!)^{a_q} \right]^{-1}
\sum_{\{k\}}{}'s_{k_1}^{n_1}s_{k_2}^{n_2}\cdots s_{k_m}^{n_m}\,.
\label{DefSummation}
\ee
In (\ref{DefSummation}), $\sum_{\{k\}}{}'$ means a summation over
all possible $m$ integers $1\leq k_1,k_2,\ldots,k_m$ $\leq p$ such
that any two of them do not coincide.
$m$ integers $1\leq n_1\leq n_2\leq\cdots\leq n_m\leq 2n$
are defined so that $a_1$ of them are $1$, $a_2$ of them are $2$
and so on.
Then we have some relations
\be
\sum_{q=1}^{2n} a_q = m,\quad
\sum_{q=1}^{2n}qa_q = \sum_{i=1}^mn_i = 2n\,.
\label{NumInd}
\ee
These summation vanish in the case when $2n$ is a odd integer,
because $I_p$ should not change its value when we change
the sign of the height $s_{k_i}\rightarrow -s_{k_i}$.
The two summations appeared in (\ref{DefR2}) are simply
described as
\ba
I_p(2)   & = & \sum_{\mbox{\small paths}}\frac{1}{2}\sum_{i=1}^ps_i^2
               \,, \nonu \\
I_p(1^2) & = & \frac{1}{2}\sum_{\mbox{\small paths}}
               \sum_{i,j}{}'s_is_j\,,
\ea
where we omit entries whose $a_i$ are $0$.
Thus all we have to do is describe $I_p(2)$ and $I_p(1^2)$ for general
$p$.

Two generating functions give us a key to solve this problem.
To our surprise, one of these generating functions is a $q$-deformed
combination or a $q$-deformed binomial coefficient.
The definition of it is similar to an ordinary combination
%
%
\be
\prod_{s=-p}^p \left( 1+\e^{sx}y \right) \equiv
\sum_{r=0}^{2p+1}{}_{2p+1}C_r(x)y^r\,,
\ee
where $p$ is a non-negative half integer.
It is easy to check that this function ${}_{2p+1}C_r(x)$ is
obtained by replacing integers with
$q$-deformed integers ($q=\exp(x/2)$)
in the usual expression for binomial coefficients,
\be
{}_{2p+1}C_r(x) = \frac{N_{2p+1}(x)!}{N_r(x)!N_{2p+1-r}(x)!}\,,
\label{DefqComb}
\ee
\[
N_r(x) = \sinh\frac{rx}{2}/\sinh\frac{x}{2}\,,
\]
\[
N_r(x)! = N_1(x)\cdot N_2(x)\cdots N_r(x)\,.
\]
We give some characteristics of the $q$-deformed combination,
\begin{eqnarray*}
{}_{2p+1}C_r(0) & = & {}_{2p+1}C_r\,, \\
\sum_{r=0}^{2p+1}{}_{2p+1}C_r(x) & = & 2^{2p+1}\prod_{r=-p}^p
                                       \cosh\frac{rx}{2}\,,
\\
\sum_{r=0}^{2p+1}(-1)^r{}_{2p+1}C_r(x) & = & 0\,.
\end{eqnarray*}
As explained in appendix A, ${}_{2p+1}C_p(x)$ is nothing but
a generating function of a following summation over the paths,
\be
\sum_{\mbox{\small paths}}\prod_{i=1}^p\e^{s_ix} =
{}_{2p+1}C_p(x) = \frac{N_{2p+1}(x)!}{N_p(x)!N_{p+1}(x)!}\,.
\label{qComb}
\ee
Now we call the expansion coefficient of $x^{2n}$
in the $q$-deformed combination $G_p(2n)$,
\[
{}_{2p+1}C_p(x) = \sum_{n=0}^\infty G_p(2n) x^{2n}\,.
\]
This coefficient $G_p(2n)$ is the sum of $I_p$ over all
partitions of $2n$,
\ba
G_p(2n)         & = & \sum_{\mbox{\small paths}}
                      \frac{1}{(2n)!}
                      \left(
                       \sum_{i=1}^p s_i
                      \right)^{2n} \nonu \\
                & = & \sum_{\mbox{\small all partitions of}\ 2n}
                      I_p(1^{a_1}\cdots (2n)^{a_{2n}})\,.
\ea
Thus $G_p(2)$ is the sum of $I_p(2)$ and $I_p(1^2)$,
\ba
G_p(2)
& = &
\sum_{\mbox{\small paths}}\frac{1}{2}
\left( \sum_{i=1}^ps_i \right)^2 \nonu \\
& = &
I_p(2) + I_p(1^2)\,.
\label{SquareOfSum}
\ea
Since Taylor expansion of a $q$-deformed integer is
\be
N_r(x) = r\left[
           1+\frac{1}{24}(r^2-1)x^2
           +\frac{1}{5760}(r^2-1)(3r^2-7)x^4 + {\cal{O}}(x^6)
          \right]\,,
\label{ExqInt}
\ee
we can compute the summation over paths in (\ref{SquareOfSum})
\ba
G_p(2) & = &
\left(
 \begin{array}{c}
  2p+1 \\ p
 \end{array}
\right)
\left(
 \sum_{r=1}^{2p+1}-\sum_{r=1}^p-\sum_{r=1}^{p+1}
\right)
\frac{1}{24}(r^2-1) \nonu \\
& = &
\frac{1}{12}(p+1)^2p
\left(
 \begin{array}{c}
  2p+1 \\ p
 \end{array}
\right)\,.
\ea
We can check this result for small $p$.

As we find the sum $I_p(2)+I_p(1^2)$, we must find one of them,
say $I_p(2)$.
To do this we define a generating function of $I_p(2n)$
\ba
\Gamma_p(x) & \equiv & \sum_{\mbox{\small paths}} \sum_{i=1}^p
                       \e^{s_ix} \nonu \\
& = &
\sum_{n=0}^\infty I_p(2n)x^{2n}\,.
\ea
There are no odd terms in $x$ due to the symmetry
$s_i\leftrightarrow -s_i$.
In appendix B we show that $\Gamma_p(x)$ is given by a summation
of $q$-deformed integers weighted by a binomial coefficient
\ba
\Gamma_p(x) & = &
\sum_{i=0}^{p-1}\sum_{j=0}^{p+1}
\left(
 \begin{array}{c}
  p+1+i-j \\ i
 \end{array}
\right)
\left(
 \begin{array}{c}
  p-1-i+j \\ j
 \end{array}
\right)
\e^{(p-i-j)x} \nonu \\
& = &
\sum_{n=0}^{p-1}
\left(
 \begin{array}{c}
  2p+1 \\ n
 \end{array}
\right)
N_{2p+1-2n}(x)\,.
\label{GammaP}
\ea
The proof of this relation is interesting.
It is hard to understand this relation algebraically.
We prove it in appendix B diagrammatically.
Comparing the coefficient of $x^2$ in both sides of (\ref{GammaP}),
we find $I_p(2)$ (see appendix B for details of calculation),
\ba
I_p(2) & = & \frac{1}{24}\sum_{n=0}^{p-1}
\left(
 \begin{array}{c}
  2p+1 \\ n
 \end{array}
\right)
(2p+1-2n)\{(2p+1-2n)^2-1\} \nonu \\
& = & \frac{1}{6}(p+1)p
\left(
 \begin{array}{c}
  2p+1 \\ p
 \end{array}
\right)\,.
\ea
Thus the rest $I_p(1^2)$ is
\be
I_p(1^2) = G_p(2) - I_p(2) =
\frac{1}{12}(p+1)p(p-1)
\left(
 \begin{array}{c}
  2p+1 \\ p
 \end{array}
\right)\,.
\ee
This vanishes when $p=1$ as expected.

Since both of $I_p(2)$ and $I_p(1^2)$ have a factor $(p+1)p$ and
a combination ${}_{2p+1}C_p$,
we can rewrite the recursion relation for $r_2$ (\ref{DefR2}) with
respect to derivatives of $W(r_0)$
\be
r_2W'(r_0)+\frac{1}{6}r_0^{(2)}r_0W''(r_0)
+\frac{1}{12}(r_0^{(1)})^2r_0W'''(r_0) = 0\,.
\ee
We already know that $r_0^{(1)}=[W'(r_0)]^{-1}$.
As for $r_0^{(2)}$,
we rewrite it with derivatives of $W(r_0)$
differentiating both sides of $x=W(r_0)$ twice with respective to
$x$,
\be
r_0^{(2)} = -W''(r_0)[W'(r_0)]^{-3}\,.
\ee
At last we succeed in describing $r_2/r_0$ by $W(r_0)$,
\ba
\frac{r_2}{r_0} & = &
\frac{1}{W'(r_0)}
\left[
  \frac{1}{6} \frac{\{W''(r_0)\}^2}{\{W'(r_0)\}^3}
 -\frac{1}{12}\frac{W'''(r_0)}{\{W'(r_0)\}^2}
\right] \nonu \\
& = &
-\frac{1}{12}
 \left( \frac{1}{W'(r_0)}\frac{d}{dr_0}\right)^2
 \log W'(r_0)\,.
\label{R2R0}
\ea
Thus we can explicitly perform the integration in the first term
of the right hand side of (\ref{BIZTorus})
\ba
\int_0^1\!\!dx(1-x)\frac{r_2}{r_0} & = &
-\frac{1}{12}\int_0^{a^2}\!\!dr(1-W(r))
\frac{d}{dr}\frac{1}{W'(r)} \frac{d}{dr} \log W'(r)
\nonu \\
& = &
\frac{1}{12}W''(0)-\frac{1}{12}\log W'(a^2)\,.
\ea
Collecting all contributions, we get to the final answer
\ba
e^{(1)} & = &
\frac{1}{12}W''(0)-\frac{1}{12}\log W'(a^2)-\frac{1}{8}W''(0)
\nonu \\
&   & -\frac{1}{12}\log a^2 + \frac{1}{24}W''(0)
\nonu \\
& = &
-\frac{1}{12}\log\{a^2W'(a^2)\}\,.
\label{E1}
\ea
The result is surprisingly simple.
A matter of special mention here is cancellation of terms which
include $W''(0)$.
We can easily check our result by a simple example of \cite{BIZ}
\be
W(r) = r+12\lambda_1r^2\,,
\ee
where $\lambda_1$ is a coefficient of $\tr \Phi^4$ in the action.
Since
\be
a^2W'(a^2) = 2-a^2\,,
\ee
we have the same answer as BIZ
\be
\e^{(1)} = -\frac{1}{12}\log(2-a^2)\,.
\label{BIZExTorus}
\ee
It is important that our result is quite general
and we can easily understand the generating function
for each model using (\ref{E1}).

If we adopt $W(r)$ discussed in section 2
\be
W(r) = 1 + \sum_{i=-1}^m
       \left(\frac{r}{k}\right)^{M_i+2}
       \left(1-\frac{r}{k}\right)^{n_i}\tau_i\,,
\ee
we can roughly understand the scaling characteristic of the toric
generating function.
Using $W(a^2)=1$ we have
\ba
\frac{a^2}{k}W'(a^2)
& = &
\left(1-\frac{a^2}{k}\right)^{k-1}
\sum_{i=-1}^m\left(\frac{a^2}{k}\right)^{M_i+2}
\left(1-\frac{a^2}{k}\right)^{n_i-k}
\nonu \\
&   &
\times
\left\{
  \frac{M_i+n_i+2-k}{k}\left(1-\frac{a^2}{k}\right)
 -\frac{n_i}{k}
\right\} \tau_i\,.
\ea
Therefore,
\ba
e^{(1)}
& = & -\frac{1}{12}\log k
      -\frac{k-1}{12}\log\left(1-\frac{a^2}{k}\right)
-\frac{1}{12}\log
\left[
 1+\sum_{i=0}^m \left(\frac{a^2}{k}\right)^{M_i+2}
 \left(1-\frac{a^2}{k}\right)^{n_i-k}
\right. \nonu \\
&   &
\times
\left.
 \left\{
  \frac{M_i+n_i+2-k}{k}\left(1-\frac{a^2}{k}\right)-\frac{n_i}{k}
 \right\} \tau_i
\right]\,.
\label{ScalingTorus}
\ea
Indeed $e^(1)$ has a logarithmic singularity with respect to
$1-a^2/k$.
If one wants to see the scaling nature of the generating
function more explicitly, a scaling unit like the cosmological
constant must be introduced.
We will argue this point in the next section.

%
\subsection{Double Torus}

Next we derive the generating function of correlation functions
for double torus.
The process of finding the answer in the double toric case
gives us a key for generalization to higher genus case.

The starting point for the generating function of double torus
is more complex than the toric case
and is read from (\ref{GeneralGF}),
\ba
e^{(2)} & = & \int_0^1\!\!dx(1-x)
              \left[
               \frac{r_4}{r_0}
               -\frac{1}{2}\left(\frac{r_2}{r_0}\right)^2
              \right] \nonu \\
&   &
-\frac{1}{192}
\left[
 22\{W''(0)\}^3-20W''(0)W'''(0)+3W^{(4)}(0)
\right] \nonu \\
&   &
\left.
 +\frac{1}{12}
 \left\{
  (1-x)\frac{r_2}{r_0}
 \right\}^{(1)}
\right|_0^1
\left.
 -\frac{1}{720}
 \left\{
  (1-x)\log\frac{r_0}{x}
 \right\}^{(3)}
\right|_0^1\,.
\label{BIZDoubleTorus}
\ea
We already know all but $r_4$.
Before we express $r_4$ by $W(r_0)$, we rewrite last two terms of
the right hand side of (\ref{BIZDoubleTorus}).
As we see in the toric case we should treat $r_2$ in the form of
$\tilde{r}_2=r_2/r_0$,
\be
\left.
 \frac{1}{12}
 \left\{
  (1-x)\frac{r_2}{r_0}
 \right\}^{(1)}
\right|_0^1
=
\left.
 -\frac{1}{12}\frac{r_2}{r_0}
\right|_0^1
\left.
 -\frac{1}{12}\left(\frac{r_2}{r_0}\right)^{(1)}
\right|_{x=0}\,.
\ee
If we use the following relation
\be
\frac{d}{dx} = r_0^{(1)}\frac{d}{dr_0} = [W'(r_0)]^{-1}\frac{d}{dr_0}
\,,
\ee
and the result for $r_2/r_0$ in the previous subsection,
we have
\be
\left(\frac{r_2}{r_0}\right)^{(1)} =
-\frac{8\{W''(r_0)\}^3-7W'(r_0)W''(r_0)W'''(r_0)
       +\{W'(r_0)\}^2W^{(4)}(r_0)}{12\{W'(r_0)\}^6}
\,.
\ee
Since $W'(0)=1$, we have
\ba
\lefteqn{
 \left.
  \frac{1}{12}
  \left\{
   (1-x)\frac{r_2}{r_0}
  \right\}^{(1)}
 \right|_0^1
} \nonu \\
& = &
-\frac{1}{144}
\left[
 2\{W''(a^2)\}^2-W'(a^2)W'''(a^2)
\right]
[W'(a^2)]^{-4} \nonu \\
&   &
+\frac{1}{144}\left[2\{W''(0)\}^2-W'''(0)\right]
\nonu \\
&   &
+\frac{1}{144}\left[8\{W''(0)\}^3-7W''(0)W'''(0)+W^{(4)}(0)\right]\,.
\ea
On the other hand, we find
\ba
\lefteqn{
 -\frac{1}{720}
 \left\{
  (1-x)\log\frac{r_0}{x}
 \right\}^{(3)}
} \nonu \\
& = &
\frac{1}{720}
\left[
 3\left\{
   -\frac{(r_0^{(1)})^2}{r_0^2}+\frac{r_0^{(2)}}{r_0}+\frac{1}{x^2}
  \right\}
 -(1-x)
 \left\{
  \frac{2(r_0^{(1)})^3}{r_0^3}-\frac{3r_0^{(1)}r_0^{(2)}}{r_0^2}
  +\frac{r_0^{(3)}}{r_0}-\frac{2}{x^3}
 \right\}
\right] \nonu \\
& = &
\frac{1}{720}
\left[
 3\left(
   -\frac{1}{r_0^2\{W'(r_0)\}^2}-\frac{W''(r_0)}{r_0\{W'(r_0)\}^3}
   +\frac{1}{\{W(r_0)-W(0)\}^2}
  \right)
\right. \nonu \\
&   &
 -(1-W(r_0))
 \left(
   \frac{2}{r_0^3\{W'(r_0)\}^3}+\frac{3W''(r_0)}{r_0^2\{W'(r_0)\}^4}
 \right.
\nonu \\
&   &
\left.
 \left.
  +\frac{3\{W''(r_0)\}^2-W'(r_0)W'''(r_0)}{r_0\{W'(r_0)\}^5}
  -\frac{2}{\{W(r_0)-W(0)\}^3}
 \right)
\right]\,,
\ea
where we used
\be
r_0^{(3)} =
\frac{3\{W''(r_0)\}^2-W'(r_0)W'''(r_0)}{W'(r_0)^5}\,.
\ee
Taking the limits of $r_0\rightarrow a^2$ and $r_0\rightarrow 0$,
we find
\ba
\lefteqn{
 \left.
  -\frac{1}{720}
  \left\{
   (1-x)\log\frac{r_0}{x}
  \right\}^{(3)}
 \right|_0^1
} \nonu \\
& = &
\frac{1}{240}
\left[
 1-\frac{W''(a^2)}{a^2\{W'(a^2)\}^3}-\frac{1}{a^4\{W'(a^2)\}^2}
\right] \nonu \\
&   &
-\frac{1}{2880}\left[ 9\{W''(0)\}^2-4W'''(0) \right] \nonu \\
&   &
-\frac{1}{2880}
\left[
 10\{W''(0)\}^3-8W''(0)W'''(0)+W^{(4)}(0)
\right]\,.
\ea

Now we rewrite $r_4$ only by $r_0$ with the help of the function
$W(r_0)$.
The recursion relation for $r_4$ is much longer than that for $r_2$.
This relation is found by identifying the coefficient of
$\epsilon^4$ in both sides of (\ref{Recr}),
\ba
0 & = &
r_4 + \sum_{p=1}^\infty 2(p+1)\lambda_p
\left[
 \left(
  \begin{array}{c}
   2p+1 \\ p
  \end{array}
 \right)
 r_4r_0^p
 +r_2\left\{
      \left(
       \begin{array}{c}
        2p+1 \\ p
       \end{array}
      \right)
      pr_2r_0^{p-1}
     \right.
\right. \nonu \\
&   &
\left.
 +\sum_{\mbox{\small paths}}\sum_i\frac{1}{2}s_i^2r_0^{(2)}r_0^{p-1}
 +\frac{1}{2}
 \sum_{\mbox{\small paths}}\sum_{i,j}{}'s_is_j
 (r_0^{(1)})^2r_0^{p-2}
\right\} \nonu \\
&   &
+r_0
\left\{
 \left(
  \begin{array}{c}
   2p+1 \\ p
  \end{array}
 \right)
 pr_4r_0^{p-1} +
 \left(
  \begin{array}{c}
   2p+1 \\ p
  \end{array}
 \right)
 \frac{1}{2}p(p-1)r_2^2r_0^{p-2}
\right. \nonu \\
&   &
+(p-1)\sum_{\mbox{\small paths}}\sum_i
 \frac{1}{2}s_i^2r_2r_0^{(2)}r_0^{p-2}
+(p-2)\frac{1}{2}\sum_{\mbox{\small paths}}\sum_{i,j}{}'s_is_j
 r_2(r_0^{(1)})^2r_0^{p-3} \nonu \\
&   &
+2\cdot\frac{1}{2}\sum_{\mbox{\small paths}}\sum_{i,j}{}'s_is_j
 r_2^{(1)}r_0^{(1)}r_0^{p-2}
+\sum_{\mbox{\small paths}}\sum_i\frac{1}{2}s_i^2
 r_2^{(2)}r_0^{p-1} \nonu \\
&   &
+\sum_{\mbox{\small paths}}\sum_i\frac{1}{4!}s_i^4r_0^{(4)}r_0^{p-1}
+\sum_{\mbox{\small paths}}\sum_{i,j}{}'\frac{1}{3!}s_is_j^3
 r_0^{(1)}r_0^{(3)}r_0^{p-2} \nonu \\
&   &
+\frac{1}{2}\sum_{\mbox{\small paths}}\sum_{i,j}{}'
 \frac{1}{4}s_i^2s_j^2
 (r_0^{(2)})^2r_0^{p-2}
+\frac{1}{2}\sum_{\mbox{\small paths}}\sum_{i,j,k}{}'\frac{1}{2}
 s_is_js_k^2(r_0^{(1)})^2r_0^{(2)}r_0^{p-3} \nonu \\
&   &
\left.
 \left.
  +\frac{1}{4!}\sum_{\mbox{\small paths}}\sum_{i,j,k,l}{}'
    s_is_js_ks_l (r_0^{(1)})^4r_0^{p-4}
 \right\}
\right]\,.
\ea
In this equation, numbers $I_p(\{a\})$ given by partitions of
two and four appear.
For $I_p(2)$ and $I_p(1^2)$, we have computed before.
Thus we have to determine 5 sums characterized by partitions
of four:
\ba
I_p(4)    & = & \sum_{\mbox{\small paths}}\sum_i
                \frac{1}{4!}s_i^4,\qquad
I_p(1\cdot 3) = \sum_{\mbox{\small paths}}\sum_{i,j}{}'
                \frac{1}{3!}s_is_j^3\,,
\nonu \\
I_p(2^2)  & = & \frac{1}{2}\sum_{\mbox{\small paths}}
                \sum_{i,j}{}'\frac{1}{4}
                s_i^2s_j^2,\qquad
I_p(1^2\cdot 2) = \frac{1}{2}\sum_{\mbox{\small paths}}\sum_{i,j,k}{}'
                \frac{1}{2}s_is_js_k^2\,,
\nonu \\
I_p(1^4)  & = & \frac{1}{4!}\sum_{\mbox{\small paths}}
                \sum_{i,j,k,l}{}' s_is_js_ks_l\,.
\ea
Applying the argument of the previous subsection, we can compute
the sum of the 5 numbers, \ie, $G_p(4)$.
Expanding the $q$-deformed combination
${}_{2p+1}C_p(x)$ up to $x^4$ order, we have
\ba
\lefteqn{
\sum_{\mbox{\small paths}}\frac{1}{4!}
\left( \sum_is_i \right)^4
} \nonu \\
& = &
G_p(4) \nonu \\
& = &
I_p(4) + I_p(1\cdot 3) + I_p(2^2)
+I_p(1^2 \cdot 2) + I_p(1^4) \nonu \\
& = &
\left[
 \frac{1}{5760}
 \left\{
  \sum_{r=1}^{2p+1}(r^2-1)(3r^2-7)+
  \left( \sum_{r=1}^p + \sum_{r=1}^{p+1} \right)
  (r^2-1)(7r^2-3)
 \right\}
\right. \nonu \\
&   &
\left.
 +\frac{1}{2} \left\{ \frac{1}{12}(p+1)^2p \right\}^2
 -\frac{1}{2} \frac{1}{24^2}
 \left(
  \sum_{r=1}^{2p+1}+\sum_{r=1}^p+\sum_{r=0}^{p+1}
 \right)
 (r^2-1)^2
\right]
\left(
 \begin{array}{c}
  2p+1 \\ p
 \end{array}
\right) \nonu \\
& = &
\frac{1}{1440}(p+1)^3p(5p^2+2p-2)
\left(
 \begin{array}{c}
  2p+1 \\ p
 \end{array}
\right)\,.
\label{SumFour}
\ea
Using the generating function $\Gamma_p(x)$ defined in
(\ref{GammaP}), we can also compute $I_p(4)$.
If we expand the $q$-deformed integers in $\Gamma_p(x)$ up to $x^4$
order, we find (see appendix B)
\ba
I_p(4) & = &
\frac{1}{5760}\sum_{n=0}^{p-1}
\left(
 \begin{array}{c}
  2p+1 \\ n
 \end{array}
\right)
(2p+1-2n)\{(2p+1-2n)^2-1\}
\nonu \\
&   &
\times \{3(2p+1-2n)^2-7\} \nonu \\
& = &
\frac{1}{360}(p+1)p(6p-1)
\left(
 \begin{array}{c}
  2p+1 \\ p
 \end{array}
\right)\,.
\label{IP4}
\ea

How we progress from now on is a problem.
The generating functions ${}_{2p+1}C_p(x)$ and $\Gamma_p(x)$
are insufficient to find all of sums given by partitions of four.
We need another 3 relations in order to decide all the unknown
quantities.
It seems that any approach using generating functions does not
work well.
It is possible to find all of these sums, however.
We can guess beforehand $p$-dependence of the unknown quantities.
If one carefully looks at the results for $I_p(2),\ I_p(1^2)$ and
$I_p(4)$ and the expansion coefficients of the $q$-deformed
combination, $G_p(2)$ and $G_p(4)$, we observe following
characteristics of them:
\begin{enumerate}
\item They always contain the combination
${}_{2p+1}C_p$
and can be divided by $(p+1)p$.
\item $G_p(2n)$ is a product of ${}_{2p+1}C_p$
and a polynomial of degree $3n$.
\item Each of $I_p(2),\ I_p(1^2)$ and $I_p(4)$ are also a product
of the combination ${}_{2p+1}C_p$ and a polynomial of $p$.
The degree of the polynomial for $I_p(1^2)$ is three, which is the
same as that for $G_p(2)$.
It is natural to think that
the degree of the polynomials are decided by the number of
summation indices.
Indeed the number of indices is both two for $G_p(2)$ and
$I_p(1^2)$.
On the other hand, the degree of the polynomial for $I_p(2)$
is less than that for $I_p(1^2)$ by one, because there is only
one summation index for $I_p(2)$.
The degree of the polynomial for $I_p(4)$ is also explained
by such a consideration.
\item Since $I_p(1^2)$ vanishes at $p=1$, it must have a factor
$(p-1)$ in the polynomial part.
In the same way, $I_p(1\cdot 3)$ and $I_p(2^2)$ should have
$(p-1)$ in the polynomial part.
There will be a factor $(p-1)(p-2)$ in the polynomial part of
$I_p(1^2\cdot 2)$ and $(p-1)(p-2)(p-3)$ for $I_p(1^4)$.
\end{enumerate}
Therefore all we have to do is decide two parameters for each of
the unknown quantities.
We assume that
\ba
I_p(1\cdot 3) & = &
(p+1)p(p-1)(a_1p+b_1)
\left(
 \begin{array}{c}
  2p+1 \\ p
 \end{array}
\right)\,, \nonu \\
I_p(2^2)      & = &
(p+1)p(p-1)(a_2p+b_2)
\left(
 \begin{array}{c}
  2p+1 \\ p
 \end{array}
\right)\,, \nonu \\
I_p(1^2\cdot 2) & = &
(p+1)p(p-1)(p-2)(a_3p+b_3)
\left(
 \begin{array}{c}
  2p+1 \\ p
 \end{array}
\right)\,.
\ea
Unknown parameters $a_1, b_1, a_2, b_2, a_3$ and $b_3$ are fixed
by explicit computations of $I_2(1\cdot 3),\ I_3(1\cdot 3)$ and so
on.
For example,
\ba
I_2(1 \cdot 3) & = &
\frac{1}{3!}[2+8+1+1-1-1+1+1+8+2] \nonu \\
& = &
\frac{11}{3}\,, \nonu \\
I_3(1 \cdot 3) & = &
\frac{1}{3!}[2\{(54+24+27+3+8+2)+(16+16+8+2+8+2) \nonu \\
&   &
\quad +2(8+2+8+2+1+1)+2(8+2) \nonu \\
&   &
\quad +(8+2-8-2-1-1)+(1+1+1+1+1+1) \nonu \\
&   &
\quad +3(1+1)+(1+1-1-1-1-1) \} + 4(-1-1)] \nonu \\
& = &
\frac{238}{3} \,.
\ea
It is a laborious work to calculate numbers such as $I_4(1^2\cdot 2)$
by hand.
Of course these quantities are easily computed if one uses
a computer.
After a long algebra, we find these quantities
\ba
I_p(1\cdot 3) & = &
\frac{1}{180}(p+1)p(p-1)(6p-1)
\left(
 \begin{array}{c}
  2p+1 \\ p
 \end{array}
\right)\,, \nonu \\
I_p(2^2)      & = &
\frac{1}{240}(p+1)p(p-1)(6p-1)
\left(
 \begin{array}{c}
  2p+1 \\ p
 \end{array}
\right)\,, \nonu \\
I_p(1^2\cdot 2) & = &
\frac{1}{360}(p+1)p(p-1)(p-2)(11p-3)
\left(
 \begin{array}{c}
  2p+1 \\ p
 \end{array}
\right)\,.
\ea
We checked that these results are valid even for other
values of $p$ which we do not use to determine the unknown
parameters.
The rest one has a factor $(p+1)p(p-1)(p-2)(p-3)$ as expected
\be
I_p(1^4) =
\frac{1}{1440}(p+1)p(p-1)(p-2)(p-3)(5p-2)
\left(
 \begin{array}{c}
  2p+1 \\ p
 \end{array}
\right)\,.
\ee
The degree of the polynomial part in $I_p(1^4)$ is six, which is
the same as that of $G_p(4)$.
These facts support validity of our assumption for the sums.

Again we can rewrite the recursion equation for $r_4$
by derivatives of $W(r_0)$,
\ba
\lefteqn{
 r_4W'(r_0)+\frac{1}{2}W''(r_0)r_2^2
 +r_2\left\{
       \frac{1}{6}W''(r_0)r_0^{(2)}
      +\frac{1}{12}W'''(r_0)(r_0^{(1)})^2
     \right\}
} \nonu \\
& + &
r_0\left\{
     \frac{1}{6}W'''(r_0)r_2r_0^{(2)}
    +\frac{1}{12}W^{(4)}(r_0)r_2(r_0^{(1)})^2
    +\frac{1}{6}W'''(r_0)r_2^{(1)}r_0^{(1)}
    +\frac{1}{6}W''(r_0)r_2^{(2)}
   \right\}
\nonu \\
& + &
 \left[
   \frac{1}{60}r_0^2W'''(r_0)
  +\frac{1}{72}r_0W''(r_0)
 \right] r_0^{(4)}
+\left[
   \frac{1}{30}r_0^2W^{(4)}(r_0)
  +\frac{11}{180}r_0W'''(r_0)
 \right] r_0^{(1)}r_0^{(3)}
\nonu \\
& + &
 \left[
   \frac{1}{40}r_0^2W^{(4)}(r_0)
  +\frac{11}{240}r_0W'''(r_0)
 \right] (r_0^{(2)})^2
\nonu \\
& + &
 \left[
   \frac{11}{360}r_0^2W^{(5)}(r_0)
  +\frac{1}{12}r_0W^{(4)}(r_0)
 \right] (r_0^{(1)})^2r_0^{(2)}
\nonu \\
& + &
 \left[
   \frac{1}{288}r_0^2W^{(6)}(r_0)
  +\frac{1}{80}r_0W^{(5)}(r_0)
 \right] (r_0^{(1)})^4
= 0\,.
\ea
We rearrange this equation using the relation found in the previous
subsection
\be
 \frac{1}{6}W''(r_0)r_0^{(2)}
+\frac{1}{12}W'''(r_0)(r_0^{(1)})^2
= -\frac{r_2}{r_0} W'(r_0)\,,
\ee
and identities
\ba
r_2^{(1)} & = &
r_0^{(1)}\frac{r_2}{r_0}
+r_0 \left( \frac{r_2}{r_0} \right)^{(1)}\,, \nonu \\
r_2^{(2)} & = &
\left[
  r_0^{(2)}\frac{r_2}{r_0}
 +2r_0^{(1)} \left( \frac{r_2}{r_0} \right)^{(1)}
\right]
+r_0 \left( \frac{r_2}{r_0} \right)^{(2)}\,,
\ea
and find
\ba
\lefteqn{
 \left[
  \frac{r_4}{r_0}-\frac{1}{2}\left( \frac{r_2}{r_0} \right)^2
 \right]
 W'(r_0)
} \nonu \\
& = &
\left[
  \frac{1}{2} \left( \frac{r_2}{r_0} \right)^2W'(r_0)
 -\frac{1}{6} \left( \frac{r_2}{r_0} \right) W''(r_0)(r_0^{(1)})^2
\right. \nonu \\
&   &
 -\frac{1}{6} \left( \frac{r_2}{r_0} \right) W''(r_0)r_0^{(2)}
 -\frac{1}{3} \left( \frac{r_2}{r_0} \right)^{(1)}W''(r_0)r_0^{(1)}
\nonu \\
&   &
-\frac{1}{72}  W''(r_0)r_0^{(4)}
-\frac{11}{180}W'''(r_0)r_0^{(1)}r_0^{(3)}
-\frac{11}{240}W'''(r_0)(r_0^{(2)})^2
\nonu \\
&   &
\left.
 -\frac{1}{12}W^{(4)}(r_0)(r_0^{(1)})^2r_0^{(2)}
 -\frac{1}{80}W^{(5)}(r_0)(r_0^{(1)})^4
\right] \nonu \\
&   &
-r_0
\left[
  \frac{1}{2} \left( \frac{r_2}{r_0} \right)^2W''(r_0)
 +\frac{1}{6} \left( \frac{r_2}{r_0} \right)^{(1)}W'''(r_0)r_0^{(1)}
 +\frac{1}{6} \left( \frac{r_2}{r_0} \right)^{(2)}W''(r_0)
\right. \nonu \\
&   &
+\frac{1}{6}  \left( \frac{r_2}{r_0} \right) W''(r_0)r_0^{(2)}
+\frac{1}{12} \left( \frac{r_2}{r_0} \right) W^{(4)}(r_0)(r_0^{(1)})^2
\nonu \\
&   &
+\frac{1}{60} W'''(r_0)r_0^{(4)}
+\frac{1}{30} W^{(4)}(r_0)r_0^{(1)}r_0^{(3)}
+\frac{1}{40} W^{(4)}(r_0)(r_0^{(2)})^2
\nonu \\
&  &
\left.
 +\frac{11}{360} W^{(5)}(r_0)(r_0^{(1)})^2r_0^{(2)}
 +\frac{1}{288}  W^{(6)}(r_0)(r_0^{(1)})^4
\right]\,.
\ea
At this stage, we have terms proportional to $r_0$ and
terms which do not depend on $r_0$ explicitly.
The relation $x=W(r_0)$ and (\ref{R2R0}) found in the toric case
allow us to write
\ba
r_0^{(4)} & = &
-\left[
  15\{W''(r_0)\}^3-10W'(r_0)W''(r_0)W'''(r_0)
 \right. \nonu \\
&   &
 \left.
  +\{W'(r_0)\}^2W^{(4)}(r_0)
 \right]
[W'(r_0)]^{-7}\,,
\ea
and
\ba
\left( \frac{r_2}{r_0} \right)^{(2)} & = &
\frac{1}{12}
\left[
  48\{W''(r_0)\}^4-59W'(r_0)\{W''(r_0)\}^2W'''(r_0)
\right. \nonu \\
&   &
 +7\{W'(r_0)\}^2\{W'''(r_0)\}^2
 +11\{W'(r_0)\}^2W''(r_0)W^{(4)}(r_0)
\nonu \\
&   &
\left.
 -\{W'(r_0)\}^3W^{(5)}(r_0)
\right]
[W'(r_0)]^{-8}\,.
\ea
We get to the following compact result
similar to that for $r_2/r_0$ in the end,
\be
\frac{r_4}{r_0}-\frac{1}{2}\left( \frac{r_2}{r_0} \right)^2 =
\left( \frac{1}{W'(r_0)}\frac{d}{dr_0} \right)^2
\left[ -r_0\Omega_1^{(2)}(r_0) + \Omega_0^{(2)}(r_0) \right]\,,
\label{R4R0}
\ee
where functions $\Omega_1^{(2)}(r_0)$ and $\Omega_0^{(2)}(r_0)$ are
described only by derivatives of $W(r_0)$,
\ba
\Omega_1^{(2)}(r_0) & = &
\frac{1}{1440}
\left[
 28\{W''(r_0)\}^3-29W'(r_0)W''(r_0)W'''(r_0)
\right. \nonu \\
&   &
\left.
 +5\{W'(r_0)\}^2W^{(4)}(r_0)
\right]
[W'(r_0)]^{-5}\,, \nonu \\
\Omega_0^{(2)}(r_0) & = &
\frac{1}{2880}
\left[
 31\{W''(r_0)\}^2-16W'(r_0)W'''(r_0)
\right]
[W'(r_0)]^{-4}\,.
\ea
Finally we have
\ba
\lefteqn{
 \int_0^1\!\!dx(1-x)
 \left[
   \frac{r_4}{r_0}
  -\frac{1}{2} \left( \frac{r_2}{r_0} \right)^2
 \right]
} \nonu \\
& = &
\int_0^{a^2}\!\!dr(1-W(r))
\frac{d}{dr}\frac{1}{W'(r)}\frac{d}{dr}
\left[ -r\Omega_1^{(2)}(r)+\Omega_0^{(2)}(r) \right]
\nonu \\
& = &
-a^2\Omega_1^{(2)}(a^2)+\Omega_0^{(2)}(a^2)
+\Omega_1^{(2)}(0)-\Omega_0^{(2)'}(0)-\Omega_0^{(2)}(0)
\nonu \\
& = &
\frac{a^2
      \left[
       28\{W''(a^2)\}^3-29W'(a^2)W''(a^2)W'''(a^2)
       +5\{W'(a^2)\}^2W^{(4)}(a^2)
      \right]}{1440\{W'(a^2)\}^5} \nonu \\
&   &
+\frac{31\{W''(a^2)\}^2-16W'(a^2)W'''(a^2)}
      {2880\{W'(a^2)\}^4} \nonu \\
&   &
+\frac{90\{W''(0)\}^3-84W''(0)W'''(0)+13W^{(4)}(0)}
      {1440}
\nonu \\
&   &
-\frac{31\{W''(0)\}^2-16W'''(0)}{2880}
\,.
\ea
As in the case of torus, terms that include $W''(0),\ W'''(0)$ and
$W^{(4)}(0)$ miraculously cancel out each other,
\begin{eqnarray*}
&   &
\frac{1}{1440}
\left[
 90\{W''(0)\}^3-84W''(0)W'''(0)+13W^{(4)}(0)
\right] \\
&   &
-\frac{1}{2880}
\left[
 31\{W''(0)\}^2-16W'''(0)
\right] \\
&   &
-\frac{1}{192}
\left[
 22\{W''(0)\}^3-20W''(0)W'''(0)+3W^{(4)}(0)
\right] \\
&   &
+\frac{1}{144}
\left[
 2\{W''(0)\}^2-W'''(0)
\right] \\
&   &
+\frac{1}{144}
\left[
 8\{W''(0)\}^3-7W''(0)W'''(0)+W^{(4)}(0)
\right] \\
&   &
-\frac{1}{2880}
\left[
 9\{W''(0)\}^2-4W'''(0)
\right] \\
&   &
-\frac{1}{2880}
\left[
 10\{W''(0)\}^3-8W''(0)W'''(0)+W^{(4)}(0)
\right] \\
&   &
= 0\,.
\end{eqnarray*}
Getting all contributions together,
we have the generating function for double torus,
\ba
e^{(2)} & = &
\frac{a^2
      \left[
       28\{W''(a^2)\}^3-29W'(a^2)W''(a^2)W'''(a^2)
       +5\{W'(a^2)\}^2W^{(4)}(a^2)
      \right]}{1440\{W'(a^2)\}^5}
\nonu \\
&   &
-\frac{9\{W''(a^2)\}^2-4W'(a^2)W'''(a^2)}
      {2880\{W'(a^2)\}^4}
\nonu \\
&   &
+\frac{1}{240}
-\frac{1}{240}\frac{W''(a^2)}{a^2\{W'(a^2)\}^3}
-\frac{1}{240}\frac{1}{a^4\{W'(a^2)\}^2}\,.
\label{E2}
\ea

Again we check our result by the example of \cite{BIZ}
\[
W(r) = r+12\lambda_1r^2\,.
\]
Since
\[
a^2W'(a^2) = 2-a^2\,,\quad a^4W''(a^2)=2(1-a^2)\,,
\]
we find
\ba
e^{(2)} & = &
 -\frac{1}{1440}\frac{28\cdot 8(1-a^2)^3}{(2-a^2)^5}
 -\frac{9}{2880}\frac{4(1-a^2)}{(2-a^2)^4} \nonu \\
& &
+\frac{1}{240}-\frac{1}{240}\frac{2(1-a^2)}{(2-a^2)^3}
-\frac{1}{240(2-a^2)^2} \nonu \\
& = &
-\frac{1}{720}
 \frac{(1-a^2)^3(82+21a^2-3a^4)}{(2-a^2)^5}\,.
\ea
This completely agrees with the result of BIZ
and guarantees validity of our method for computing summations
over the paths.
If you remember how laborious the calculation of BIZ is,
you will be satisfied with the convenience of our result
(\ref{E2}).

%
\subsection{$h\ge 3$ Cases}

We succeeded in solving $e^{(h)}$ for given $W(r_0)$ up to $h=2$.
It is possible to apply our method for computing generation
functions to higher genus case.

The most difficult point is how to compute the sums
over paths $I_p(1^{a_1}\cdots(2n)^{a_{2n}})$ given by a partition
of an even integer $2n$.
These summations are necessary for solving the recursion relation
for $\tilde{r}_{2s}$ defined in (\ref{tilder}).
Once we overcome this point, it is straightforward to solve the
generating functions for higher genus.
Of course, it is a hard work from the practical point of view.

As discussed before, $G_p(2n)$ and $I_p(2n)$ are both products of
the combination ${}_{2p+1}C_p$ and a polynomial of $p$.
Our success in double toric case encourages us to assume that
general $I_p(1^{a_1}\cdots(2n)^{a_{2n}})$ has the same structure
as $G_p(2n)$ or $I_p(2n)$,
\be
I_p(1^{a_1}\cdots(2n)^{a_{2n}}) =
A(p)
\left(
 \begin{array}{c}
  2p+1 \\ p
 \end{array}
\right)\,,
\ee
where $A(p)$ is a polynomial of $p$.
This assumption is reliable.
Let us consider the degree of the polynomial $A(p)$.
As easily confirmed by expanding ${}_{2p+1}C_p(x)$ with respective
to $x^2$, the degree of the polynomial part of $G_p(2n)$ is $3n$.
Since the number of indices of the summation in $G_p(2n)$ is $2n$
and that in general $I_p(1^{a_1}\cdots(2n)^{a_{2n}})$ is
$m$ defined by (\ref{NumInd}),
the degree of $A(p)$ is $3n-2n+m=n+m$.
Furthermore we know that $I_p(1^{a_1}\cdots(2n)^{a_{2n}})$ vanishes
for $p=1,2,\ldots,m-1$ and $A(p)$ always has a factor $(p+1)p$.
Indeed the difference of $G_p(2n)$ and $I_p(2n)$ has a factor
$(p-1)$, since ${}_3C_1(x)=\Gamma_1(x)=N_3(x)$.
Thus we can assume the detailed form of the polynomial part
\ba
\lefteqn{I_p(1^{a_1}\cdots(2n)^{a_{2n}})} \nonu \\
& = &
\bar{A}(p)(p+1)p(p-1)\cdots(p-m+1)
\left(
 \begin{array}{c}
  2p+1 \\ p
 \end{array}
\right)\,,
\label{IP}
\ea
where $\bar{A}(p)$ is a polynomial of degree $n-1$.
Since this polynomial is determined by fixing $n$ coefficients,
we need $n$ individual values for
$I_p(1^{a_1}\cdots(2n)^{a_{2n}}),\ p=m,m+1,\ldots,m+n-1$
to determine each of the sum completely.
We can express $r_{2h}$ or $\tilde{r}_{2h}$ with respect to
derivatives of $W(r_0)$ using the recursive equation for
$r_{2h}$ and the knowledge for $\tilde{r}_{2s},\ (0\leq s<h)$.
This means that we can solve $e^{(h)}$ for any $h$ in principle.
In general,
the larger the genus is, the harder the computation of the sums
will become.
We have full information for the generating functions in exchange
for complexity of the calculation.
If one wants to compute the generating functions for $h\geq 3$,
one must rely on a computer program to understand the sums over
the paths.

Now we guess what happens when one calculates general $e^{(h)}$.
First, from our experience for toric and double toric cases,
$\tilde{r}_{2h}, \ (h\geq 1)$ will be written in a form like
\be
\tilde{r}_{2h} =
\left( \frac{1}{W'(r_0)}\frac{d}{d r_0} \right)^2
R_{2h}(r_0)\,.
\label{TildeR}
\ee
In (\ref{TildeR}) an unknown function $R_{2h}(r_0)$ will have
following structure,
\be
R_{2h}(r_0) = \sum_{s=0}^{h-1}(-1)^sr_0^s\Omega_s^{(h)}(r_0)\,,
\label{Omegash}
\ee
where $\Omega_s^{(h)}(r_0)$ are written by only derivatives
of $W(r_0)$.
This is based on the fact there is $h$ unfixed coefficients
in the polynomial part of $I_p$ given by a partition of
$2h$ as we showed in (\ref{IP}).
We have already used these notations in the double toric case.
For the toric case,
\[
\Omega_0^{(1)}=-\frac{1}{12}\log W'(r_0)\,.
\]
As we will see in section 5,
only $\Omega_{h-1}^{(h)}$ contributes in the double scaling limit.
This means that only the first term of the right hand side of
(\ref{GeneralGF}) contributes in this limit.
Probably there will be a rule in the form of $\Omega_s^{(h)}(r_0)$.
Our method cannot answer this question satisfactorily.
Our conjecture for $\tilde{r}_{2h}$ is important, for
it guarantees that we can perform integration in (\ref{GeneralGF})
explicitly.
The case of sphere is the simplest but is exceptional, for
$\tilde{r}_0$ cannot be written in the form
of (\ref{TildeR}) and there are regular terms in $e^{(0)}$.

One of supports of validity of our method is quite
a surprising cancellation in terms which include derivatives
of $W(r_0)$ at $r_0=0$.
It is natural to think that these cancellation will also happen
in the case of $h\geq 3$.
This conjecture is expected from our knowledge that
the regular part is present only in genus $0$.
This cancellation is concerned with the fact that there is a function
$E^{(h)}(r)$ such that
\[
e^{(h)} = E^{(h)}(a^2) - E^{(h)}(0)
\]
up to a constant.
We can regard $-E^{(h)}(0)$ as the regular part and $E^{(h)}(a^2)$
as the singular part.
We find such functions for $h=0,1,2$:
\ba
E^{(0)}(r) & = &
\int_k^r\!\!\frac{dr'}{r'}(1-W(r))^2\,, \nonu \\
E^{(1)}(r) & = &
-\frac{1}{12}\log \frac{rW'(r)}{W(r)}\,, \nonu \\
E^{(2)}(r) & = &
-\frac{r \left[
          28\{W''(r)\}^3-29W'(r)W''(r)W'''(r)
          +5\{W'(r)\}^2W^{(4)}(r)
         \right]}{1440\{W'(r)\}^5} \nonu \\
&   &
-\frac{9\{W''(r)\}^2-4W'(r)W'''(r)}{\{W'(r)\}^4} \nonu \\
&   &
+\frac{1}{240W(r)}
-\frac{1}{240}\frac{W''(r)}{r\{W'(r)\}^3}
-\frac{1}{240}\frac{1}{r^2\{W''(r)\}^2}\,.
\ea
One can easily check that
$-E^{(0)}(0) = e^{(0)}_{\mbox{\small bulk}},\,
  E^{(1)}(0) = E^{(2)}(0) = 0$.

%
\section{Correlation Functions and Leading Singular Terms}

In this section we make a study of correlation functions of
scaling operators in 1-matrix models.
It is well understood that these correlation functions obey
a scaling rule in the continuum theory \cite{DDK}.
Our results for the generating functions of the correlation
functions are exact.
As we do not take any continuum limit,
we can study the finite size effects of the correlation functions
as well as the universality of the models.

We must prepare a scaling unit to understand scaling properties
of the correlation functions.
Here we set up the model by the function $W(r)$ in
(\ref{GeneralW}).
Then the system is fine tuned in the $k$-th criticality and
has $(m+1)$ source terms.
We introduce the scaling unit as the $0$-th scaling operator
as announced in section 2.
It is convenient to choose a puncture operator
\be
\rho_0^0[\Phi] = \frac{1}{12k^2}\tr\Phi^4
\ee
for this purpose.
In this section, we use a letter $t$ as the scaling unit instead of
$\tau_0$.
This is nothing but the rescaled cosmological constant.
The scaling unit $t$ is slightly different from the rescaled
cosmological constant of \cite{GRMI} $\tilde{t}$.
There is a relation between these two scaling units
\[
t = N^{-2k/(2k+1)}\tilde{t}\,.
\]
In the double scaling limit, we take a limit $N\rightarrow \infty$
keeping $\tilde{t}$ finite.
This procedure indeed extracts only the universal part.
Therefore the most singular contributions with respect to $t$
are universal and survive in the double scaling limit.
The remainder are regarded as finite size corrections.

We may choose a coupling constant of other scaling operator
as a scaling unit.
However marginal scaling operators whose gravitational
dimension are one are not suitable for the scaling unit.
The reason why marginal scaling operators are not accepted is
that coupling constants of them cannot scale.
We have already met such a situation.
The fine tuned part of the action $S_{\mbox{\small cr}}$ is
effectively a marginal scaling operator and its coupling constant
is $\tau_{-1}=-1$, \ie, a number.
Such a change of the scaling unit brings a seemingly new critical
behavior.
Effective values of the critical exponent and the gravitational
dimensions of scaling operators change into new values.
In the end of this section we argue this point.

It is convenient to introduce
\be
f(t,\{\tau\}) \equiv 1-\frac{a^2}{k},\quad
W(a^2)=1\,,
\ee
as a function of the cosmological constant $t$ and $m$ coupling
constants $\{\tau\}$.
The condition $W(a^2)=1$ is translated as a condition
for $f$,
\ba
0 & = & \sum_{i=-1}^m\sum_{p=0}^{M_i+2}
\left(
 \begin{array}{c}
  M_i+2 \\ p
 \end{array}
\right)
(-1)^p f^{n_i+p}\tau_i
\nonu \\
& = &
-f^k+(1-f)^2 t+\sum_{i=1}^m\sum_{p=0}^{M_i+2}
\left(
 \begin{array}{c}
  M_i+2 \\ p
 \end{array}
\right)
(-1)^p f^{n_i+p}\tau_i \,.
\label{f}
\ea
This equation gives a unique solution for $f(t,\{\tau\})$,
if one solves (\ref{f}) iteratively.
Using the technique of the appendix A of \cite{GRMI},
the solution of (\ref{f}) is
\ba
\lefteqn{f(t,\{\tau\})} \nonu \\
&\!\! = \!\!&
\frac{1}{k}\sum_{p=1}^\infty
\frac{1}{p!}
\left( \frac{\partial}{\partial t} \right)^{p-1}
\left[
 \left\{
  \sum_{i=1}^\infty\tau_i (1-t^{1/k})^{M_i}t^{n_i/k}
  -t\left( \frac{1}{(1-t^{1/k})^2}-1 \right)
 \right\}^p
 t^{1/k-1}
\right] \nonu \\
&\!\!   \!\!&
+t^{1/k}\,.
\label{Solf}
\ea
We can understand scaling properties of correlation functions
more clearly through $f$.
As the first step we rewrite the generating functions for genus
$0$, $1$ and $2$ with respective to $f$ and its $t$-derivatives.

We begin with the case of sphere.
Since the bulk part does not contribute to scaling properties of
correlation functions, only the singular part
\be
e^{(0)}_{\mbox{\small sing}} =
-\frac{1}{2}\sum_{i=-1}^m\sum_{j=-1}^m\tau_i\tau_j
\sum_{p=0}^\infty\frac{(-1)^p}{n_i+n_j+1+p}
\left(
 \begin{array}{c}
  M_i+M_j+3 \\ p
 \end{array}
\right)
(-1)^p f^{n_i+n_j+1+p}
\ee
is necessary for us,
We can simplify this if we take the second derivative
of $e_{\mbox{\small sing}}^{(0)}$ with respect to $t$.
Since $t$ is the cosmological constant, this second derivative
stands for specific heat.
To see this we differentiate once,
\ba
\frac{\partial}{\partial t}e^{(0)}_{\mbox{\small sing}}
& = &
-\frac{1}{2}\sum_{i=-1}^m\sum_{j=-1}^m\tau_i\tau_j
\sum_{p=0}^\infty(-1)^p
\left(
 \begin{array}{c}
  M_i+M_j+3 \\ p
 \end{array}
\right)
f^{n_i+n_j+p}f^{(1)} \nonu \\
&   &
-\sum_{i=-1}^m\tau_i\sum_{p=0}^{M_i+3}
\frac{(-1)^p}{n_i+1+p}
\left(
 \begin{array}{c}
  M_i+3 \\ p
 \end{array}
\right)
f^{n_i+1+p}\,.
\ea
Here we use identities
\be
\left(
 \begin{array}{c}
  M_i+M_j+4 \\ p
 \end{array}
\right)
=
\left(
 \begin{array}{c}
  M_i+M_j+3 \\ p
 \end{array}
\right)
+
\left(
 \begin{array}{c}
  M_i+M_j+3 \\ p-1
 \end{array}
\right)
\ee
and $(W(a^2)-1)^2=0$ to derive
\ba
0 & = &
\sum_{i=-1}^m\sum_{j=-1}^m\tau_i\tau_j
\sum_{p=0}^\infty(-1)^p
\left(
 \begin{array}{c}
  M_i+M_j+4 \\ p
 \end{array}
\right)
f^{n_i+n_j+p} \nonu \\
& = &
\sum_{i=-1}^m\sum_{j=-1}^m\tau_i\tau_j
\sum_{p=0}^\infty(-1)^p
\left(
 \begin{array}{c}
  M_i+M_j+3 \\ p
 \end{array}
\right)
f^{n_i+n_j+p} (1-f)\,.
\ea
Since $f$ is not one, we have
\be
\frac{\partial}{\partial t}e^{(0)}_{\mbox{\small sing}} =
-\sum_{i=-1}^m\tau_i\sum_{p=0}^{M_i+3}
\frac{(-1)^p}{n_i+1+p}
\left(
 \begin{array}{c}
  M_i+3 \\ p
 \end{array}
\right)
f^{n_i+1+p}\,.
\ee
In the same way we differentiate again,
\ba
\frac{\partial^2}{\partial t^2}e^{(0)}_{\mbox{\small sing}}
& = &
-\sum_{i=-1}^m\tau_i\sum_{p=0}^{M_i+3}(-1)^p
\left(
 \begin{array}{c}
  M_i+3 \\ p
 \end{array}
\right)
f^{n_i+p} f^{(1)}
\nonu \\
&   &
-\sum_{p=0}^3\frac{(-1)^p}{1+p}
\left(
 \begin{array}{c}
  3 \\ p
 \end{array}
\right)
f^{1+p}
\nonu \\
& = &
-\sum_{i=-1}^m\tau_i\sum_{p=0}^{M_i+2}(-1)^p
\left(
 \begin{array}{c}
  M_i+2 \\ p
 \end{array}
\right)
f^{n_i+p} f^{(1)}(1-f) \nonu \\
&   &
-\frac{1}{4}[1-(1-f)^4]\,.
\ea
Finally we get to
\be
\frac{\partial^2}{\partial t^2}e^{(0)}_{\mbox{\small sing}} =
-\frac{1}{4}[1-(1-f)^4] =
-f \left( 1-\frac{3}{2}f+f^2-\frac{1}{4}f^3 \right)\,.
\label{SHeat}
\ee
Therefore the function $f$ is almost the specific heat.
The deviation of the right hand side of (\ref{SHeat}) from
$-f$ represents a finite size effect of the model,
since these terms disappear in the double scaling limit.
Thus $e^{(0)}_{\mbox{\small sing}}$ is given by
\be
e^{(0)}_{\mbox{\small sing}} =
F(t, \{\tau\})\,,
\label{FRep0}
\ee
where $F(t, \{\tau\})$ satisfies
\[
\frac{\partial^2}{\partial t^2}F(t, \{\tau\}) =
-\frac{1}{4}[1-(1-f)^4]\,.
\]

Next is the case of torus.
In this case, we do not have to differentiate $e^{(1)}$.
Since
\be
W(a^2) = 1 + \sum_{i=-1}^m\sum_{p=0}^{M_i+2}
\left(
 \begin{array}{c}
  M_i+2 \\ p
 \end{array}
\right)
(-1)^p
\left( 1-\frac{a^2}{k} \right)^{n_i+p}\tau_i\,,
\ee
differentiating both sides of (\ref{f}) once leads to
\ba
0 & = &
\sum_{i=-1}^m\tau_i\sum_{p=0}^{M_i+2}
\left(
 \begin{array}{c}
  M_i+2 \\ p
 \end{array}
\right)
(-1)^p(n_i+p)f^{n_i+p-1} f^{(1)} \nonu \\
&   &
+\sum_{p=0}^2
\left(
 \begin{array}{c}
  2 \\ p
 \end{array}
\right)
(-1)^pf^p \nonu \\
& = &
-k W'(a^2) f^{(1)} + (1-f)^2\,,
\ea
where $f^{(n)}\equiv\partial^n f/\partial t^n$.
We have
\be
kW'(a^2) = \frac{(1-f)^2}{f^{(1)}}
\ee
and
\ba
a^2 W'(a^2) & = &
\frac{a^2}{k}\cdot kW'(a^2) \nonu \\
& = &
\frac{(1-f)^3}{f^{(1)}}\,.
\ea
Therefore $e^{(1)}$ is written as
\ba
e^{(1)} & = &
-\frac{1}{12}\log \{a^2 W'(a^2)\} \nonu \\
& = &
\frac{1}{12}\log f^{(1)} -\frac{1}{4}\log (1-f)\,.
\label{FRep1}
\ea
In the end we rewrite $e^{(2)}$ by $f$.
In the same way as we derive $kW'(a^2)$ we find
\be
k^2W''(a^2) =
\frac{(1-f)^2f^{(2)}}{(f^{(1)})^3} + \frac{4(1-f)}{f^{(1)}}\,,
\ee
\ba
k^3W'''(a^2) & = &
\frac{(1-f)^2[3(f^{(2)})^2-f^{(1)}f^{(3)}]}{(f^{(1)})^5} \nonu \\
&   &
+ \frac{6(1-f) f^{(2)}}{(f^{(1)})^3} + \frac{6}{f^{(1)}}\,,
\ea
\ba
k^4W^{(4)}(a^2) & = &
\frac{(1-f)^2[15(f^{(2)})^3-10f^{(1)}f^{(2)}f^{(3)}
     +(f^{(1)})^2f^{(4)}]}{(f^{(1)})^7} \nonu \\
&   &
+\frac{8(1-f)[3(f^{(2)})^2-f^{(1)}f^{(3)}]}{(f^{(1)})^5}
+\frac{12 f^{(2)}}{(f^{(1)})^3}\,.
\ea
If we note $\bar{W}^{(n)} \equiv k^nW^{(n)}(a^2)$,
\ba
e^{(2)} & = &
-\frac{(1-f)[28(\bar{W}^{(2)})^3
            -29\bar{W}^{(1)}\bar{W}^{(2)}\bar{W}^{(3)}
            +5(\bar{W}^{(1)})^2\bar{W}^{(4)}]}
      {1440(\bar{W}^{(1)})^5}
\nonu \\
&    &
-\frac{9(\bar{W}^{(2)})^2-4\bar{W}^{(1)}\bar{W}^{(3)}}
      {1880(\bar{W}^{(1)})^4}
\nonu \\
&   &
+\frac{1}{240} - \frac{\bar{W}^{(2)}}{240(1-f)(\bar{W}^{(1)})^3}
-\frac{1}{240(1-f)^2(\bar{W}^{(1)})^2}
\nonu \\
& = &
-\frac{16(f^{(2)})^3-21f^{(1)}f^{(2)}f^{(3)}+5(f^{(1)})^2f^{(4)}}
      {1440(f^{(1)})^4(1-f)^3}
\nonu \\
&   &
+\frac{215(f^{(2)})^2-156f^{(1)}f^{(3)}}
      {2880(f^{(1)})^2(1-f)^4}
\nonu \\
&   &
-\frac{47f^{(2)}}{120(1-f)^5}
-\frac{593(f^{(1)})^2}{720(1-f)^6}
+\frac{1}{240}\,.
\label{FRep2}
\ea
This result is a little complex.
We will be able to rewrite the generation functions for $h\geq 3$
in this manner and
the results will be more complicated than the double toric case.


Eqs.(\ref{FRep0}), (\ref{FRep1}), (\ref{FRep2}) and the formal solution
for $f$ (\ref{Solf}) give us exact answers for the correlation
functions of scaling operators,
\be
\langle\rho_{n_1/k}^{M_1}\cdots\rho_{n_p/k}^{M_p}\rangle^{(h)} \equiv
\frac{\partial}{\partial\tau_1}\cdots
\frac{\partial}{\partial\tau_p}
\left. e^{(h)} \right|_{\{\tau\}=0} \,.
\ee
Here we do not go into details, for we cannot find any useful
things in finite size corrections.
We will go deep into the leading contributions soon.
If one asks for exactness, we must study influence of
the deformation parameter $b$ discussed in section 2.
If we define
\[
f(t, \{\tau\}, b) \equiv 1 - \frac{a^2}{k+b}\,,
\]
Eqs.(\ref{FRep0}), (\ref{FRep1}) and (\ref{FRep2}) hold good
as they are, while the equation for $f$ is a little modified to
\be
0 =
-f^k+b(1-f)f^k+(1-f)^2t
+\sum_{i=1}^m\sum_{p=0}^{M_i+2}
\left(
 \begin{array}{c}
  M_i+2 \\ p
 \end{array}
\right)
(-1)^pf^{n_i+p}\tau_i\,.
\ee


We can extract leading singular contributions from the exact
generating functions.
These contributions survives in the double scaling limit.
{}From now on we add a subscript $0$ to indicate the most singular
contribution and take $m=\infty$.
The equation for $f_0(t,\{\tau\})$ can be easily read from (\ref{f})
by discarding $p\neq 0$ terms,
\be
\sum_{i=-1}^\infty f_0^{n_i}\tau_i = 0\,,
\ee
or
\be
t = f_0^k - \sum_{i=1}^\infty\tau_i f_0^{n_i}\,.
\label{f0}
\ee
This equation is the same as the one for specific heat
given by Gross and Migdal (GM) \cite{GRMI} when they studied matrix
models on the sphere.
Indeed we can easily understand that $f_0$ is the specific heat,
if we pick up the leading contribution from (\ref{SHeat}),
\be
\frac{\partial^2}{\partial t^2}e^{(0)}_{0,\mbox{\small sing}} = -f_0
\,.
\ee
Therefore the generating function for sphere is
\be
e^{(0)}_{0,\mbox{\small sing}} = -F_0(t, \{\tau\})\,,
\ee
where $F_0$ is related to $f_0$ as
$(\partial^2/\partial t^2)F_0 = f_0$.
Since our result for $e^{(0)}_{0,\mbox{\small sing}}$ is
exactly the same as that of GM, the correlation functions
of scaling operators also coincide with their results.
We do not repeat any longer.

It is important to pay attention to absence of the labels $M$ in
(\ref{f0}).
This means that all representations $\rho^M_{n/k}[\Phi]$ with
different $M$ but the same $n$ are degenerate in the double scaling
limit.
This shows that the indices $M$ are non-universal while $n$ is
universal.
Therefore we omit this non-universal index $M$ here
and use $\rho_{n/k}[\Phi]$ for
the scaling operator with dimension $n/k$ in this approximation.
The $M$-dependence is associated with the finite size effect
of matrix models.
If one wants to study finite size effects,
one must learn much about the $M$ dependence of the generating
functions.

Let us consider an influence of the deformation parameter $b$
which connects a criticality with a neighboring one.
Formally only the equation for $f$ is modified when $b\neq 0$.
According to this modification, the equation for $f_0$ is modified
to a form as
\be
t = (1-b)f_0^k - \sum_{i=1}^\infty\tau_i f_0^{n_i}\,,
\label{Modifiedf0}
\ee
where $b\neq 1$.
This shows that the modified equation (\ref{Modifiedf0}) is
essentially equivalent to the original one (\ref{f0})
by simultaneous rescaling of $t$ and $\{\tau\}$.
Therefore this deformation parameter is non-universal
if we do not fine tune it.

The greatest advantage of our method is that we can compute
the leading singular terms of the correlation functions
in higher genus case.
Such correlation functions have been formally studied
in the double scaling limit \cite{GRMI}.
It was, however, a hard job to compute them explicitly.
Our method offers an easier procedure for computing the
correlation functions for higher genus than the method
using the double scaling limit.

{}From (\ref{FRep1}) and (\ref{FRep2}) we can write down
the leading part of the generating functions immediately,
\ba
e^{(1)}_0 & = &
\frac{1}{12}\log f^{(1)}_0\,, \nonu \\
e^{(2)}_0 & = &
-\frac{16(f^{(2)}_0)^3-21 f^{(1)}_0f^{(2)}_0f^{(3)}_0
       +5(f^{(1)}_0)^2f^{(4)}_0}
      {1440(f^{(1)}_0)^4}\,.
\label{E01E02}
\ea
If one traces terms in the leading contributions to their origin,
they come from $\Omega_{h-1}^{(h)}$ as mentioned in the previous
section.
Anyway one can pick up the leading terms roughly taking
$f\sim t^{1/k}$.
As explained in the appendix A of \cite{GRMI}, we have
\be
f_0(t, \{\tau\}) =
\frac{1}{k}\sum_{p=1}^\infty\frac{1}{p!}
\left( \frac{\partial}{\partial t} \right)^{p-1}
\left[
 \left( \sum_{i=1}^\infty \tau_i t^{n_i/k} \right)^p t^{1/k-1}
\right]
+t^{1/k}\,.
\label{Exf0}
\ee
Thus the derivatives of $f_0$ are $(n\ge 1)$
\be
f^{(n)}_0 =%
\frac{1}{k}\sum_{p=0}^\infty\frac{1}{p!}
\left( \frac{\partial}{\partial t} \right)^{p+n-1}
\left[
 \left( \sum_{i=1}^\infty \tau_i t^{n_i/k} \right)^p t^{1/k-1}
\right]\,.
\label{Exf0n}
\ee
The leading contribution for the correlation functions of scaling
operators are defined as
\be
\langle\rho_{n_1/k}\cdots\rho_{n_p/k}\rangle_0^{(h)}\equiv
\frac{\partial}{\partial\tau_1}\cdots
\frac{\partial}{\partial\tau_p}
\left. e_0^{(h)} \right|_{\{\tau\}=0}\,.
\ee
Using (\ref{E01E02}), (\ref{Exf0}) and (\ref{Exf0n}), we find
the free energies, one-point functions and two-point functions
for torus and double torus:
%
\be
\left. e_0^{(1)} \right|_{\{\tau\}=0} =
-\frac{1}{12}\log k
-\frac{k-1}{12k}\log t\,,
\label{FE1}
\ee
%
%
\ba
\langle \rho_{n_1/k} \rangle^{(1)}_0 & = &
\frac{n_1+1-k}{12k}t^{n_1/k-1}\,, \nonu \\
\langle \rho_{n_1/k}\rho_{n_2/k} \rangle^{(1)}_0 & = &
\frac{1}{12k^2}
[(n_1+n_2+1-k)(n_1+n_2+1-2k)
\nonu \\
&   &
-(n_1+1-k)(n_2+1-k)]
t^{(n_1+n_2)/k-2}\,,
\ea
%
%
\be
\left. e^{(2)}_0 \right|_{\{\tau\}=0} =
\frac{(k-1)(2k+3)}{720k}t^{-2-1/k}\,,
\label{FE2}
\ee
%
%
\ba
\langle \rho_{n_1/k} \rangle^{(2)}_0 &\!\!\! = \!\!\!&
-\frac{1}{1440k^3}(n_1+1-k)
[5(n_1+1-2k)(n_1+1-3k)(n_1+1-4k) \nonu \\
&\!\!\!   \!\!\!&
-21(n_1+1-2k)(n_1+1-3k)(1-k)
+ 3(n_1+1-2k)(1-k)(9-2k) \nonu \\
&\!\!\!   \!\!\!&
-(1-k)(11+11k-2k^2)]t^{n_1/k-3-1/k}\,,
\ea
%
%
\ba
\lefteqn{\langle \rho_{n_1/k}\rho_{n_2/k} \rangle_0^{(2)}}
\nonu \\
& = &
-\frac{1}{1440k^3}t^{(n_1+n_2)/k-4-1/k}\times \nonu \\
&   &
[
 5(n_1+n_2+1-k)(n_1+n_2+1-2k)(n_1+n_2+1-3k)
  (n_1+n_2+1-4k)
\nonu \\
&   &
\times (n_1+n_2+1-5k)
\nonu \\
&   &
- 21(n_1+n_2+1-k)(n_1+n_2+1-2k)(n_1+n_2+1-3k)
\nonu \\
&   &
  \times (n_1+n_2+1-4k)(1-k)
\nonu \\
&   &
+  3(n_1+n_2+1-k)(n_1+n_2+1-2k)(n_1+n_2+1-3k)(1-k)(9-2k)
\nonu \\
&   &
-   (n_1+n_2+1-k)(n_1+n_2+1-2k)(1-k)(11+11k-2k^2)
\nonu \\
&   &
- 10(n_1+1-k)(n_1+1-2k)(n_1+1-3k)(n_1+1-4k)(n_2+1-k)
\nonu \\
&   &
- 21(n_1+1-k)(n_1+1-2k)(n_1+1-3k)(n_2+1-k)(n_2+1-2k)
\nonu \\
&   &
- 21(n_1+1-k)(n_1+1-2k)(n_2+1-k)(n_2+1-2k)(n_2+1-3k)
\nonu \\
&   &
- 10(n_1+1-k)(n_2+1-k)(n_2+1-2k)(n_2+1-3k)(n_2+1-4k)
\nonu \\
&   &
+ 63(n_1+1-k)(n_1+1-2k)(n_1+1-3k)(n_2+1-k)(1-k)
\nonu \\
&   &
+ 96(n_1+1-k)(n_1+1-2k)(n_2+1-k)(n_2+1-2k)(1-k)
\nonu \\
&   &
+ 63(n_1+1-k)(n_2+1-k)(n_2+1-2k)(n_2+1-3k)(1-k)
\nonu \\
&  &
-  3(n_1+1-k)(n_1+1-2k)(n_2+1-k)(1-k)(43-22k)
\nonu \\
&  &
-  3(n_1+1-k)(n_2+1-k)(n_2+1-2k)(1-k)(43-22k)
\nonu \\
&   &
+  2(n_1+1-k)(n_2+1-k)(1-k)(49-17k-2k^2)
]\,.
\ea
Especially our results for free energies
(\ref{FE1}) and (\ref{FE2}) are consistent to
the asymptotic genus expansion of the whole specific heat given
in the appendix C of \cite{GRMI}.
If we sum up these free energies like (\ref{TopEx}),
the sum obeys the same nonlinear differential equations given in
\cite{GRMI}.
Moreover we can easily read the leading contribution $e_0^{(h)}$
from the exact generating function $e^{(h)}$.
Once $e^{(h)}$ is computed by means of our method,
we can understand all about correlation functions of scaling
operators in the double scaling limit.
It is important that we do not take any continuum limit here.
We can reproduce the asymptotic topological expansion of the
correlation functions in the continuum limit when $N=1$.
This expansion converges even when the cosmological constant
$t$ is small, if we take the matrix size $N$ sufficiently large.


We close this section by arguing other choices for the scaling
unit.
Let us choose a scaling operator $\rho_{n/k}^M,\ (n\neq k)$ in place
of the puncture operator $\rho_0^0$.
Again we assign a letter $t$ for the scaling unit instead of $\tau_0$.
Then the equation for $f$ is
\be
0 =
-f^k+(1-f)^{M+2}f^n t+\sum_{i=1}^\infty\sum_{p=0}^{M_i+2}
\left(
 \begin{array}{c}
  M_i+2 \\ p
 \end{array}
\right)
(-1)^p f^{n_i+p}\tau_i \,.
\label{Modf}
\ee
In the same way as we find $f$-dependence of the specific heat,
we have
\be
\frac{\partial^2}{\partial t^2}e_{\mbox{\small sing}}^{(0)} =
-\int_0^f\!\!d\bar{f}\bar{f}^{2n}(1-\bar{f})^{2M+3}\,.
\label{ModSH}
\ee
Here we discuss only the spherical case for simplicity.
When $n=M=0$, (\ref{ModSH}) returns to (\ref{SHeat})

To study scaling properties of this system,
we pick out leading contributions.
The equation (\ref{Modf}) is simplified to
\[
0=-f_0^k+f_0^nt+\sum_{i=0}^\infty f_0^{n_i}\tau_i\,.
\]
The solution for $f_0$ is
\ba
f_0(t, \{\tau\})
& = &
\frac{1}{k-n}\sum_{p=1}^\infty\frac{1}{p!}
\left( \frac{\partial}{\partial t} \right)^{p-1}
\left[
 \left( \sum_{i=1}^\infty \tau_i t^{(n_i-n)/(k-n)} \right)^p
 t^{1/(k-n)-1}
\right] \nonu \\
&   &
+t^{1/(k-n)}\,.
\label{ModAns}
\ea
This small $t$ expansion is valid only for $k>n$.
When $n>k$, we must use $t^{-1}$ as a scaling unit.
Eq.(\ref{ModAns}) indicates that the scaling operator
$\rho_{n_i/k}^{M_i}$ acquires a new gravitational dimension
\[
\bar{d}_i = \frac{n_i-n}{|k-n|} = \frac{d_i-d_0}{|1-d_0|}\,,
\]
where $d_i=n_i/k,\,d_0=n/k$.
It is important that the dimension of the $0$-th scaling operator
and that of the marginal scaling operator are always zero and one
respectively.
The new dimension is allowed to have a negative value.
We can find another critical exponent like the string susceptibility
from the specific heat (\ref{ModSH}).
Extracting the most singular contribution from (\ref{ModSH}),
\be
\frac{\partial^2}{\partial t^2}e_{0,\mbox{\small sing}}^{(0)} =
-\frac{f_0^{2n+1}}{2n+1}\,.
\ee
We can define a new exponent $\bar{\gamma}$ in the same way as the
usual one,
\be
\left.
 \frac{\partial^2}{\partial t^2}e_{0,\mbox{\small sing}}^{(0)}
\right|_{\{\tau\}=0}
\propto
\left\{
 \begin{array}{ll}
  t^{-\bar{\gamma}} & (\mbox{for}\quad n<k)\,, \\
  t^{ \bar{\gamma}} & (\mbox{for}\quad n>k)\,.
 \end{array}
\right.
\label{ModStr}
\ee
{}From (\ref{ModAns}) and (\ref{ModStr}) we have
\be
\bar{\gamma} = -\frac{2n+1}{|k-n|}
             = \frac{\gamma-2d_0}{|1-d_0|}\,.
\ee
Thus the new exponent is always negative as the $n=0$ case.
The scaling behavior of the correlation functions are characterized
by the new string susceptibility $\bar{\gamma}$ and
the new scaling dimensions $\bar{d}_i$ of sources.
It is natural that there is a universality similar to the original
one even when we change the scaling unit.

%
\section{Discussion}

We have solved general 1-matrix models keeping the matrix size $N$
finite.
It is wonderful that these random lattice models are exactly
solvable.
Our explicit answers for the generating functions
offers many information for the discreteness of the models
as well as the universality.
Especially, our method is fit for studying the correlation
functions of scaling operators.
The structure of the bulk terms found in section 2 is the most
important result which comes from the discreteness of the models.
In our approach, to take the double scaling limit is equivalent to
selecting leading singular terms as shown in section 5.
Then is it necessary for us to take the double scaling limit
in order to interpret 1-matrix models as two dimensional quantum
gravity?
In the double scaling limit, all finite size corrections are
ignored and the general covariance is exactly restored.
However, if we require the restoration of the general covariance
only approximately and allow extremely small finite size effects,
the topological expansion for small cosmological constant is
possible when the matrix size $N$ is sufficiently large.
In this point of view, this approximate universality
is a consequence of the randomness of the theory.
Since the large-$N$ expansion is a topological expansion, we can
expect the spherical topology is dominant.
This is desirable as a theory of gravity except for approximate
restoration of general covariance.
Nevertheless we need the results in the double scaling limit,
for our method is not fit for deriving the hierarchies of the
nonlinear differential equations for the specific heat function.
Moreover an interpretation of matrix models as non-critical string
theories attracts us.
If we can understand the specific heat functions of 1-matrix models
as systematic as those in the double scaling limit,
we may not have to take this special continuum limit.

In our method, there is a mathematically inexact point.
Of course, validity of our method is guaranteed by many supporting
evidences.
The point is the assumption for the $p$-dependence of sums over
the paths (\ref{IP}).
Though we convince this assumption is quite right,
a mathematically exact proof is necessary.
We must study the mathematical characteristics of the sums
more deeply.

Since we solve the recursion relation for $r_\epsilon$ (\ref{Recr}),
we can find $e^{(h)}$ only one by one.
It is important to integrate our exact results for $e^{(h)}$
and find equations like nonlinear differential equations
for the specific heat function in the double scaling limit.
It might be impossible, because hierarchies of such differential
equations are closely related to universality of matrix models.
The existence of such equations will make our understanding for
matrix models rich all the same.

Though we can solve 1-matrix models exactly, it is nontrivial
whether we can extend our approach to 2-matrix models.
It is very difficult to introduce discrete version of the scaling
operators systematically even in the Ising case.
We have no clue to solve this problem.

%
%
\par
\begin{flushleft}
\bf Acknowledgement\\[2mm]
\end{flushleft}
This work is supported in part by Grant-in-Aid for Scientific Research from
the Ministry of Education, Science and Culture.

\setcounter{section}{0}
\setcounter{equation}{0}
%
%
\appendix{$q$-deformed Combination}

In this appendix we prove (\ref{qComb}).
For the reader's information, we give the first few generating
functions:
\begin{eqnarray*}
{}_3C_1(x) & = & \e^{-x}+1+\e^x\,, \\
{}_5C_2(x) & = & \e^{-3x}+\e^{-2x}+2\e^{-x}+2+2\e^x+\e^{2x}+\e^{3x}
                 \,, \\
{}_7C_3(x) & = & \e^{-6x}+\e^{-5x}+2\e^{-4x}+3\e^{-3x}+4\e^{-2x}+
                 4\e^{-x}+5 \\
           &   & +4\e^x+4\e^{2x}+3\e^{3x}+2\e^{4x}+\e^{5x}+\e^{6x}
                 \,.
\end{eqnarray*}
First of all, we rewrite the definition of the $q$-deformed
combination (\ref{DefqComb}) introducing
$y = \e^{(p+1)x}z,\ \sigma=s+p+1$,
\ba
\prod_{s=-p}^p\left( 1+\e^{sx}y \right)
& = &
\prod_{\sigma=1}^{2p+1}\left( 1+\e^{\sigma x}z \right) \nonu \\
& \equiv &
\sum_{r=0}^{2p+1}{}_{2p+1}\tilde{C}_r(x) z^r\,.
\ea
One of the new coefficients is also described as a sum over the
paths,
\be
{}_{2p+1}\tilde{C}_p(x) =
\sum_{\mbox{\small paths}}\prod_{i=1}^p\e^{\sigma_i x} =
{}_{2p+1}C_p(x)\e^{p(p+1)x}\,,
\ee
where $\{\sigma_i | \sigma_i = 1,2,\ldots,2p+1\}$ is another set of
$p$ integers and $\sigma_i$ is equal to the number of steps until
$i$-th descending down plus one.
For the path drawn in Fig.1 of section 1, they are
$\{\sigma_1,\sigma_2,\sigma_3,\sigma_4,\sigma_5\}=
 \{1, 3, 4, 8, 11\}$.
A relation between $s_i$ and $\sigma_i$,
\[
s_i=\sigma_i-2i,\ (i=1,2,\ldots,p)\,
\]
completes the proof of (\ref{qComb}),
\ba
\sum_{\mbox{\small paths}}\prod_{i=1}^p\e^{s_i x}
& = &
\left(
 \sum_{\mbox{\small paths}}\prod_{i=1}^pe^{\sigma_i x}
\right)
\e^{-p(p+1)x} \nonu \\
& = &
{}_{2p+1}C_p(x)\,.
\ea

%
%
\appendix{Generating Function for $I_p(2n)$}

Here we prove (\ref{GammaP}) and explain how to compute
$I_p(2)$ and $I_p(4)$.
In contrast to the proof of (\ref{qComb}) in appendix A,
the proof of (\ref{GammaP}) is diagrammatic.
For reference, we give the first few of $\Gamma_p(x)$:
\begin{eqnarray*}
\Gamma_1(x) &\!\! = \!\!& \e^{-x}+1+\e^x\,, \\
\Gamma_2(x) &\!\! = \!\!& \e^{-2x}+6\e^{-x}+6+6\e^x+\e^{2x}\,, \\
\Gamma_3(x) &\!\! = \!\!& \e^{-3x}+8\e^{-2x}+29\e^{-x}+29
                          +29\e^x+8\e^{2x}+\e^{3x}\,.
\end{eqnarray*}

We explain the first equality in (\ref{GammaP}) using Fig.2.
There are $(p+1)p$ descending segments, which are identified by
a set of integers $(i,j)$ indicated in the figure.
Let us consider a descending segment CD labeled by $(i,j)$.
Since the height of the point C is $p-i-j$,
the contributions to $\Gamma_p(x)$ from the segment CD
is the product of $\e^{(p-i-j)x}$ and the number of all paths
which pass through the segment CD.
On the other hand the number of such paths equals to the product of
the number of paths from A to C and the number of paths from D to B.
These numbers correspond to the two binomial coefficients in
(\ref{GammaP}).
As the generating function $\Gamma_p(x)$ is the sum
over all contributions from $(p+1)p$ descending segments,
the first equality in (\ref{GammaP}) holds.
%
%
\begin{center}
\leavevmode
\epsfysize=10cm \epsfbox{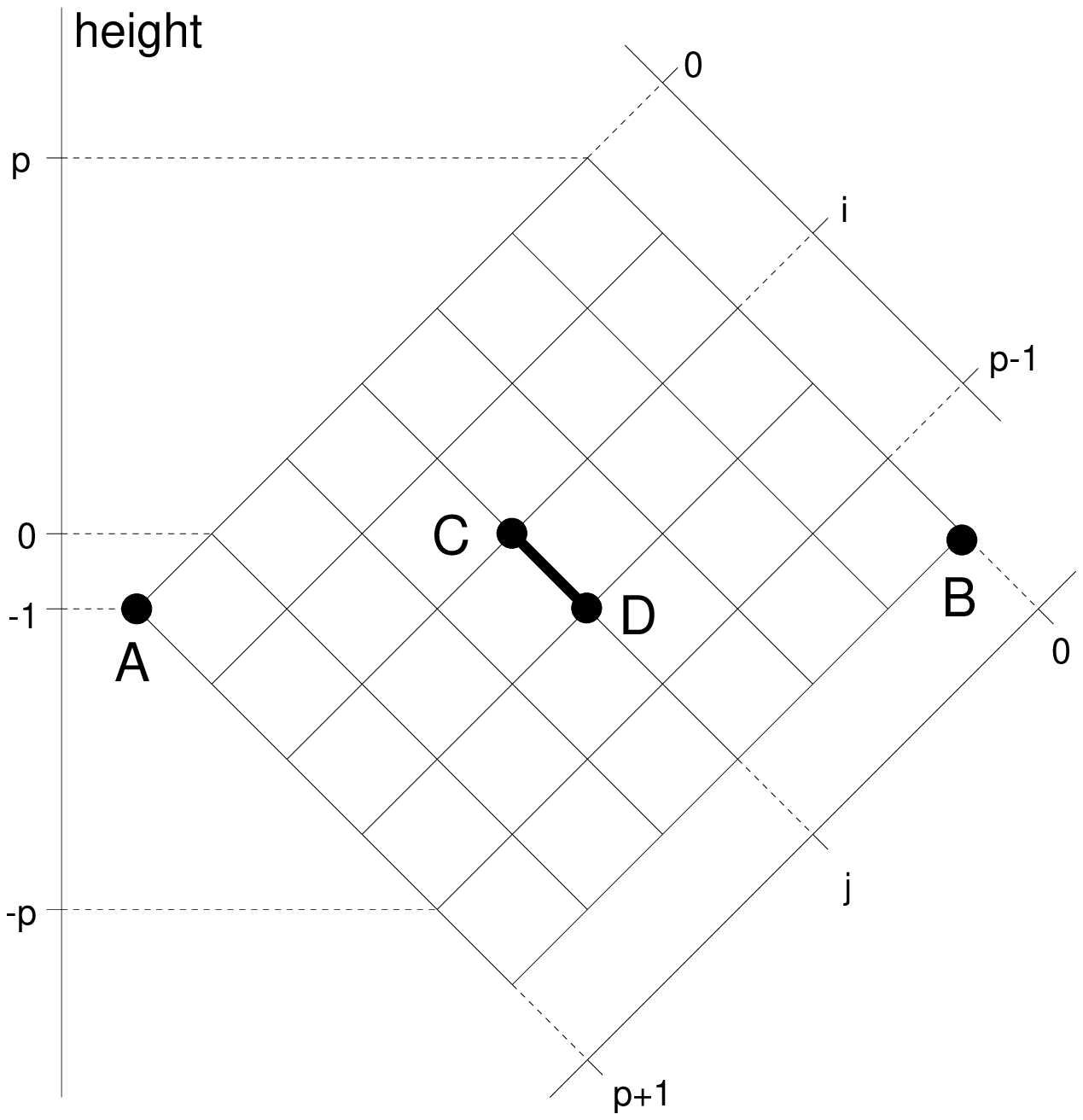}
\end{center}
\begin{center}
{\small Fig.2}
\end{center}

To prove the second equality in (\ref{GammaP}), we need a device.
Comparing the coefficients of $\e^{(p-m)x}$ in both sides of
the last equality of (\ref{GammaP}), the proof resolves itself into
the proof of
\ba
\sum_{n=0}^m
\left(
 \begin{array}{c}
  2p+1 \\ n
 \end{array}
\right)
& = &
\sum_{i=0}^m
\left(
 \begin{array}{c}
  p+1-m+2i \\ i
 \end{array}
\right)
\left(
 \begin{array}{c}
  p-1+m-2i \\ m-i
 \end{array}
\right)\,,%
\label{Fold}
\\
\sum_{n=0}^{p-1}
\left(
 \begin{array}{c}
  2p+1 \\ n
 \end{array}
\right)
& = &
\sum_{i=0}^{p-1}
\left(
 \begin{array}{c}
  1+2i \\ i
 \end{array}
\right)
\left(
 \begin{array}{c}
  2p-1-2i \\ p-i
 \end{array}
\right)\,,
\label{Fold1}
\ea
where $m=0,1,\ldots,p-1$.
The case when $m=p$ is special and we must separate as (\ref{Fold1}).
As we discussed above, the right hand side of (\ref{Fold}) describes
the number of the paths which pass through the $(m+1)$ segments at
height $p-m$ marked by a circle (see Fig.3(a)).
Here the paths which pass two marked segments are counted twice.
In the same way the paths which pass $l,\,(=1,2,\ldots,m+1)$
marked segments are counted $l$ times.
We turn back each of such paths on the line of height $p-m$
(the broken line of Fig.3(a)) fixing the part from the point A to
the point such that the path cuts the broken line first.
%
%
\begin{center}
\leavevmode
\epsfysize=7cm \epsfbox{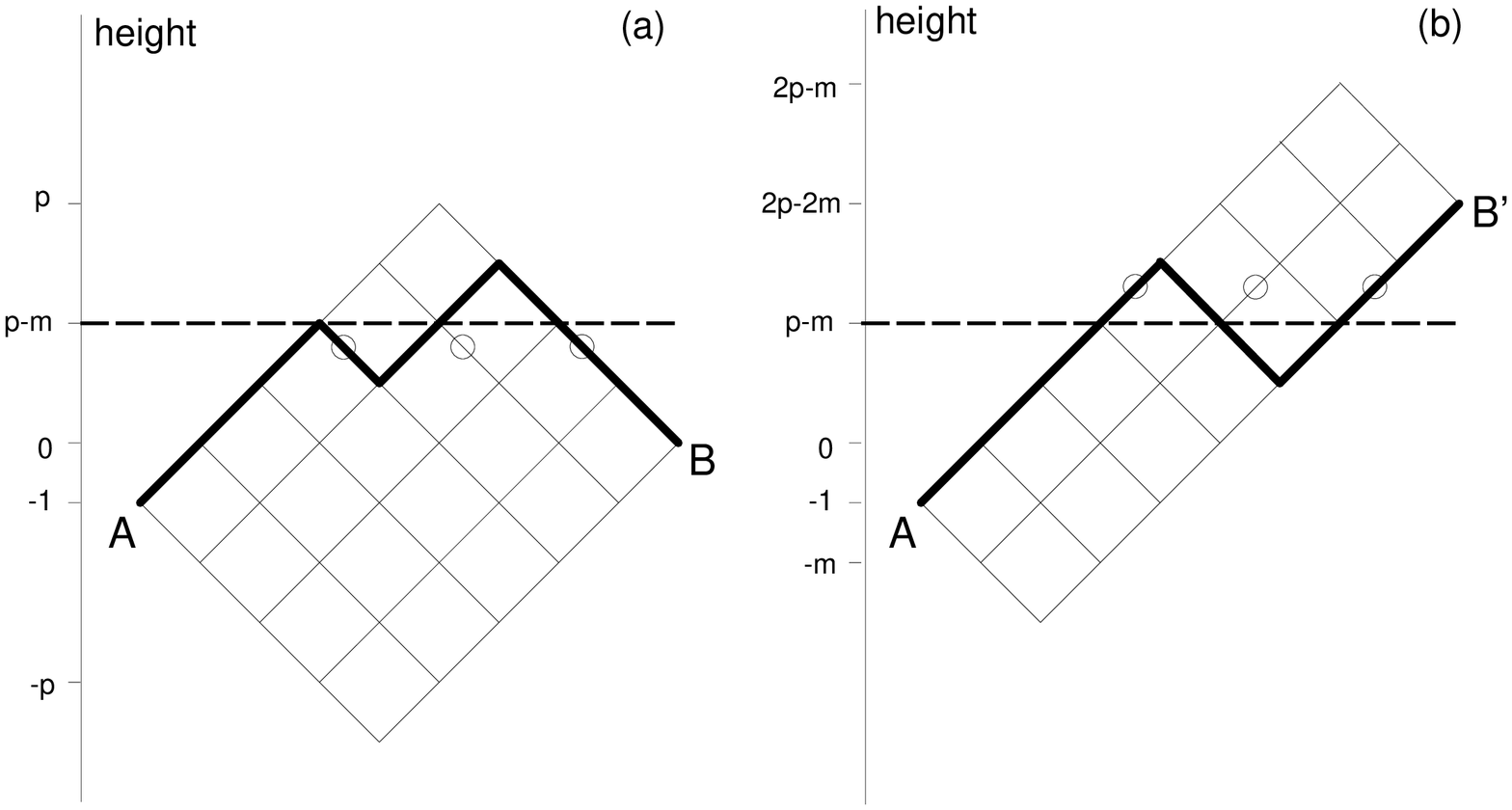}
\end{center}
\begin{center}
{\parbox{13cm}{
\small
Fig.3: A path which passes through at least one of the marked
segments is described by a thick line in Fig.3(a).
This path is `folded' on the broken line of height $p-m$
and turns to the path described by a thick line in Fig.3(b).
After this folding procedure, all $(m+1)$ marked segments
locate just above the broken line.
The marked segments which the path passes does not change
before and after this folding procedure.
In this example, $p=4$ and $m=2$.
}}
\end{center}

\noindent
Each of the `folded' path corresponds to each of the paths
which connects A with $\mbox{B}'$ as illustrated in Fig.3(b).
The marked segments are transformed into the marked ones in
Fig.3(b).
Since the number of paths from A to $\mbox{B}'$ is
${}_{2p+1}C_m$, the number of the paths in Fig.3(a) which pass
through the marked segment at least once is also ${}_{2p+1}C_m$.
This is nothing but the contribution in the left hand side of
(\ref{Fold}) from $n=m$.
If one notices that there is a overcounting in the right hand side
of (\ref{Fold}), it is easy to guess that the number of the paths
which pass through the marked segment at least $l$ times is
${}_{2p+1}C_{m+1-l}$.
If this is true, we can prove (\ref{Fold}) taking
$n=m+1-l,\,(n=0,1,\ldots,m)$.
This guess is easily justified by folding each of paths in the same
way as before moreover $2(m-n)=2(l-1)$ times shifting up the height
of line for folding by one.
Each of the resultant paths corresponds to each of the paths
which connects diagonal points of a lattice of size
$n\times(2p+1-n)$.
We can prove (\ref{Fold1}) in a similar way.
In this case, each of paths which passes through the marked segments
at least $l$ times must be turned back $2l$ times ($l=1,2,\ldots,p$)
shifting the height of line for folding by one.

Next we compute $I_p(2)$,
\[
I_p(2) = \frac{1}{24}\sum_{n=0}^{p-1}
\left(
 \begin{array}{c}
  2p+1 \\ n
 \end{array}
\right)
(2p+1-2n)\{(2p+1-2n)^2-1\}\,.
\]
We put $P\equiv 2p+1$ and fix polynomials
$a(P),b(P),c(P)$ and $d(P)$ defined by
\ba
(P-2n)^3-(P-2n) & \equiv &
  a(P)(P-n)(P-1-n)(P-2-n) \nonu \\
&   &
+ b(P)(P-n)(P-1-n) \nonu \\
&   &
+ c(P)(P-n) + d(P)\,.
\ea
If we substitute $n=P$, we get
\[
d(P) = -P(P^2-1)\,.
\]
In similar way, we find
\[
c(P) = 6(P-1)^2,\quad
b(P) = -12(P-2),\quad
a(P) = 8\,.
\]
If one uses identities like
\[
\left(
 \begin{array}{c}
  2p+1 \\ n
 \end{array}
\right)
(2p+1-n)(2p-n)(2p-1-n) =
(2p+1)(2p)(2p-1)
\left(
 \begin{array}{c}
  2p-2 \\ n
 \end{array}
\right)\,,
\]
the quantity of interest is
\ba
I_p(2) & = & \frac{1}{24}
\left[
 8(2p+1)(2p)(2p-1)\sum_{n=0}^{p-1}
 \left(
  \begin{array}{c}
   2p-2 \\ n
  \end{array}
 \right)
\right. \nonu \\
&   &
-12(2p-1)(2p+1)(2p)\sum_{n=0}^{p-1}
 \left(
  \begin{array}{c}
   2p-1 \\ n
  \end{array}
 \right)
\nonu \\
&   &
+6(2p)^2(2p+1)\sum_{n=0}^{p-1}
 \left(
  \begin{array}{c}
   2p \\ n
  \end{array}
 \right)
\nonu \\
&   &
\left.
 -(2p+2)(2p+1)(2p)\sum_{n=0}^{p-1}
  \left(
   \begin{array}{c}
    2p+1 \\ n
   \end{array}
  \right)
\right]
\nonu \\
& = & \frac{1}{24}
\left[
 4(2p+1)(2p)(2p-1)
 \left\{
  2^{2p-2}+
  \left(
   \begin{array}{c}
    2p-2 \\ p-1
   \end{array}
  \right)
 \right\}
\right. \nonu \\
&   &
-6(2p-1)(2p+1)(2p)\times 2^{2p-1}
\nonu \\
&   &
+3(2p+1)(2p)^2
 \left\{
  2^{2p}-
  \left(
   \begin{array}{c}
    2p \\ p
   \end{array}
  \right)
 \right\}
\nonu \\
&   &
\left.
 -\frac{1}{2}(2p+2)(2p+1)(2p)
  \left\{
   2^{2p+1}-
   2 \left(
      \begin{array}{c}
       2p+1 \\ p
      \end{array}
     \right)
  \right\}
\right]
\nonu \\
& = &
\frac{1}{24}(2p+1)(2p)
\left[
 4(2p-1)
 \left(
  \begin{array}{c}
   2p-2 \\ p-1
  \end{array}
 \right)
 -6p
 \left(
  \begin{array}{c}
   2p \\ p
  \end{array}
 \right)
\right. \nonu \\
&   &
\left.
 +(2p+2)
 \left(
  \begin{array}{c}
   2p+1 \\ p
  \end{array}
 \right)
\right]\,.
\ea
It is interesting that all terms which have a power of two cancel
each other.
Thus we have
\ba
I_p(2)
& = &
\frac{1}{6}(p+1)p
\left(
 \begin{array}{c}
  2p+1 \\ p
 \end{array}
\right)\,.
\ea
The computation of $I_p(4)$ is performed in the same way.
The cancellation appeared in $I_p(2)$ case also occurs
when one computes $I_p(4)$.
It seems to a common characteristic in computing $I_p(2s)$.
Of course we can calculate $I_p(2s)$ by the parameter
fitting discussed in \S 4.3.
Probably this parameter fitting will be easier than the method of
this appendix.

\vspace{3mm}
%
%

%
\end{document}